\documentclass[aps,pre,amsmath,amssymb,twocolumn,groupedaddress]{revtex4-2}

\usepackage{hyperref}

\usepackage{graphicx}
\usepackage{dcolumn}
\usepackage{bm}
\usepackage{xcolor}

\begin{document}


\title{Nonequilibrium dynamics and entropy production of a trapped colloidal particle in a complex nonreciprocal medium}
 
\author{Lea Fernandez}
\author{Siegfried Hess}
\email{s.hess@campus.tu-berlin.de}
\author{Sabine H. L. Klapp}
\email{sabine.klapp@tu-berlin.de}
\affiliation{%
Institut für Theoretische Physik, Technische Universität Berlin, Hardenbergstr. 36, D-10623 Berlin, Germany\\
}

\date{\today}

\begin{abstract}
We discuss the two-dimensional motion of a Brownian particle that is confined to a harmonic trap and driven by a shear flow. The surrounding medium induces memory effects modelled by a linear, typically nonreciprocal coupling of the particle coordinates to an auxiliary (hidden) variable. The system's behavior resulting from the microscopic Langevin equations for the three variables 
is analyzed by means of exact moment equations derived from the Fokker-Planck representation, and numerical Brownian Dynamics (BD) simulations.
Increasing the shear rate beyond a critical value we observe, for suitable coupling scenarios with nonreciprocal elements, a transition from a stationary to an instationary state, corresponding to an escape from the trap. We analyze this behavior, analytically and numerically,
in terms of the associated moments of the probability distribution, and from the perspective of nonequilibrium thermodynamics. Intriguingly, the entropy production rate remains finite when crossing the stability threshold.
\end{abstract}

\maketitle
\newpage
\section{\label{sec:intro}Introduction}
In recent years, the investigation of nonreciprocal couplings in physical systems
has received strong interest. Nonreciprocal coupling arise in many areas of physics, from macroscopic predator-prey systems \cite{volterra_1926_lotka_volterra_model,lotka_1920_lotka_volterra_model,mobilia_taeuber_2007_phase_transitions_lotka-volterra},
motile agents (e.g., birds) with restricted vision cones \cite{barberis_peruani_2016_minimal_cognitive_flocking_model,barberis_peruani_2016_minimal_cognitive_flocking_model},
(active) colloids in nonequilbrium, e.g. chemotactic, environments \cite{saha_scalar_active_mixtures_2020,meredith2020predator}, complex plasmas \cite{Ivlev_2015_statistical_mechanics_where_newtons_third_law_is_broken},
to quantum-optical systems with non-hermitian equations of motion \cite{metelmann_2015_nonreciprocal_photon_transmission}.
From the classical side, much research has recently been devoted to the collective behavior of nonreciprocal many-particle systems, such as the emergence of time-dependent states 
\cite{You_Baskaran_Marchetti_2020_pnas,fruchart_2021_nonreciprocal_phase_transitions} and of long-range ordering \cite{loos2023long}.

However, nonreciprocity can occur already on the one-particle level. Examples are a colloidal particle under time-delayed feedback control \cite{loos2019fokker}, 
an active (e.g., Ornstein-Uhlenbeck) particle with fluctuating propulsion speed \cite{Dabelow2019}, and a colloidal particle coupled to different heat baths leading to a minimal heat engine (Brownian gyrator)
\cite{filliger2007brownian,cerasoli2018asymmetry}. Importantly, nonreciprocity can also occur when modelling the non-Markovian dynamics of a tracer in a complex (e.g., viscoelastic) medium with the Markovian embedding technique \cite{siegle2010markovian}, that is, by introduction of auxiliary variables \cite{berner2018oscillating,jung2018generalized,Doerries_2021}.
In many cases, the nonreciprocity apparent in the equations of motion leads not only to interesting nonequilibrium dynamics, but also has intriguing thermodynamic consequences \cite{Loos_2020}.

In the present paper we investigate the two-dimensional Brownian motion of an (overdamped) colloidal particle under a harmonic confinement, a typical set-up to model an optical tweezer (see, e.g., \cite{franosch2011resonances}).
The resulting linear restoring force is
supplemented by non-conservative, and nonreciprocal, forces from two sources: first, we assume that the particle is subject to a linear shear flow, introducing an external drive with anisotropic characteristics. 
This situation, that is, a (trapped) Brownian particle under shear, has already been studied before (see, e.g., \cite{hess1984diffusion,hess2003crossover,holzer2010dynamics,vezirov2013nonequilibrium,PhysRevE.107.064102}).
As a new ingredient, we couple the two positional coordinates linearly and \emph{nonreciprocally} to an auxiliary (hidden) variable. 
These couplings cannot be derived from a Hamiltonian (and thus, they violate Newton's third law). Physically, we consider the auxiliary variable as a simple strategy 
to introduce memory effects as if the particle was immersed in a complex medium (see \cite{Doerries_2021,berner2018oscillating,ginot2022barrier,jung2018generalized,kappler2019non} for similar approaches).
A more general analysis with many auxiliary variables can be found elsewhere \cite{AAPPAAPP.1011A3}.
In addition to these deterministic forces, the particle is subject to friction and (white) noise.
The resulting equations of motion are linear and Markovian, and therefore fully accessible to analytical calculations. Here we mainly work using the Fokker-Planck (Smoluchowski) representation, that is, we focus on ensemble averages.
In particular, we analyze the dynamics of the driven, trapped particle based on exact equations of motion for the positional moments of the probability distribution. 

Besides exploring the dynamical behavior itself (that was partially discussed already in \cite{AAPPAAPP.1011A3}), we also analyze the system from the perspective of nonequilibrium thermodynamics. There is, indeed, a large body of literature on thermodynamic notions for linear Langevin system within and out of equilibrium \cite{Crisanti2012,puglisi2009irreversible,Loos_2020,netz2018fluctuation,Netz_2020_interacting_many-particle_systems,Doerries_2021,plati2023thermodynamic},
including, e.g., discussions of dissipation rates and fluctuation-dissipation relations, relations to information, and thermodynamic bounds \cite{plati2023thermodynamic}. Many of these studies rely on concepts from stochastic thermodynamics \cite{sekimoto2010stochastic,seifert2012stochastic}.
In the present study, we focus on (contributions to) the ensemble-averaged entropy production, considering this quantity as a measure of the distance from equilibrium. 
While some of the derived expressions are not new from a conceptual perspective, we here provide explicit analytical expressions for a physical motivated parameter choice.
Furthermore, our expressions include the rarely studied case of boundaries, and we also give relations to mechanical properties. 
Our analytical results are verified by numerical results from the solution of the underlying Langevin equations, i.e., Brownian Dynamics (BD) simulations.

We focus on the case of uniform temperatures and friction constants for all variables. 
As a main result, we show that due to the nonreciprocal coupling to the auxiliary variable, the stationary state of the shear-driven, trapped particle, i.e., localized motion, becomes unstable:
the particle escapes from the trap.
This instability occurs at a critical shear rate $\Gamma_c$.
On the Fokker-Planck level, the instability is clearly revealed by a divergence of the quadratic moments. To handle this situation within the BD simulations, we have proposed 
a control mechanism based on a source-sink set-up \cite{AAPPAAPP.1011A3} that we here formulate also on the Fokker-Planck level.
Having explored the dynamical behavior, we then characterize the different regimes below and above $\Gamma_c$ via the (total) entropy production rate. As expected, this quantity vanishes in the absence of shear and any auxiliary variable (i.e., memory). In contrast, nonreciprocal coupling to an auxiliary variable leads a finite entropy production even at $\Gamma=0$.
Increasing $\Gamma$ from zero in the completely nonreciprocal case, our analytical expressions predict, for $\Gamma<\Gamma_c$, changes of the entropy production rate in full agreement with numerical 
simulations. Interestingly the entropy production rate stays finite even across the instability. This behavior is consistent with what is seen in controlled BD simulations.

The remainder of the paper is organized as follows.
In Sec.~\ref{sec:model_eqm} we present our model and its solution in a generic form, that is, without specifying (yet) the coupling constants.
We start in Secs.~\ref{sec:langevin} and \ref{sec:fokker} with the Langevin- and Fokker-Planck representation of the dynamics. 
Sec.~\ref{sec:source} is devoted to the BD simulations, in particular, the source-sink set-up which leads to additional terms within the Fokker-Planck description.
In Sec.~\ref{sec:averages} we then present exact equations for the dynamics of averages.

In Sec.~\ref{sec:specific_model} we adapt our model to the situation of interest, that is, a trapped particle in a linear Couette flow subject to nonreciprocal couplings to an auxiliary variable.
We provide explicit expressions for averages as well as stability conditions.

Sec.~\ref{sec:thermo} is devoted to mechanical and thermodynamic properties. We start in Sec.~\ref{sec:mechanics} with derivations of mechanical quantities, particularly the angular momentum. 
The latter plays a key role for the ensemble-averaged entropy production rate discussed in Sec.~\ref{sec:entropy}. We there provide general expressions for the ensemble-averaged entropy production rate 
and explicit results for systems without and with coupling to an auxiliary variable.

Results for specific parameter choices are given in Sec.~\ref{sec:results}. 
We first discuss in detail the dynamics of the first and second positional moments below and across the stability threshold. As a second step, we consider the entropy production rate.
Finally, we present in Sec.~\ref{sec:concl} some conclusions and an outlook. The paper contains several appendices with additional analytical expressions, including a non-Markovian representation of the equations of motion.


\section{Model and equations of motion}
\label{sec:model_eqm}
\subsection{\label{sec:langevin}Langevin equations}
We consider the two-dimensional, overdamped motion of a colloidal particle with position vector $\mathbf{r}(t)= (x(t),y(t))$
in a thermal bath. Besides thermal fluctuations and friction, the particle is subject to several types of deterministic forces. These include, first, linear confining forces for each physical variable $x$, $y$, thereby mimicking a particle in
a harmonic optical trap. Second, the components of the resulting two-dimensional oscillator are mutually coupled by a volume-conserving shear flow, such as a planar Couette flow.
Third, we assume couplings to an auxiliary variable which is exposed to its own thermal bath (and restoring force). 
With the auxiliary (hidden) variable, we model in a simplistic manner the presence of a complex, viscoelastic medium introducing (exponential) memory \cite{Doerries_2021,berner2018oscillating,jung2018generalized,kappler2019non}.
The general case involving more than two physical degrees of freedom and 
$n_{\text{int}}>1$ auxiliary variables is discussed in \cite{AAPPAAPP.1011A3}.

For a compact notation including the auxiliary variable, we introduce the $n=3$-dimensional vector $\mathbf{x}(t)$ with components $x_i$, $i=1,2,3$, where $x_1$ and $x_2$ refer to $x$ and $y$, respectively,
and $x_3$ stands for the auxiliary variable. Although our focus is on the case $n=3$, we also provide,  where appropriate, general expressions for arbitrary $n$.

All deterministic forces related to $\mathbf{x}(t)$ are contained in the vector 
\begin{align}
\label{eq:vector_F}
\mathbf{F}= -\mathbf{a}\cdot\mathbf{x}+\mathbf{M}\cdot \mathbf{x},
\end{align}
where the quantities $\mathbf{a}$ and $\mathbf{M}$  are $n \times n$ matrices (i.e., second-rank tensors).
The first term in Eq.~(\ref{eq:vector_F}) models the restoring "spring" forces; thus, $\mathbf{a}$ is diagonal with positive diagonal elements  $a^{(i)}>0$. The remaining force ingredients (shear flow, coupling to auxiliary variable) are contained in the second term where $\mathbf{M}$ is assumed to have vanishing diagonal elements. Note that, in general, $\mathbf{M}$ is not symmetric, and thus, the couplings are nonreciprocal.
A specification of the elements of $\mathbf{M}$ is given in
Sec.~\ref{sec:specific_model}.

In addition to the vector notation in Eq.~(\ref{eq:vector_F}) we frequently make use of its component form,
\begin{eqnarray}
 \label{eq:Fi}   
F_i   &=&  - a^{(i)} x_i   +  M_{i j}  x_j\nonumber\\
& = & K_{ij}x_j,
 \end{eqnarray} 
 where $M_{ii}=0$, and the summation convention is used for subscripts (i.e., $M_{i j} x_j  $ stands for 
$\sum_{j = 1}^n M_{i j} x_j $). With the second line, we have introduced the elements $K_{ij}=-a^{(i)}\delta_{ij}+M_{ij}$ 
of the matrix $\mathbf{K}=-\mathbf{a}+\mathbf{M}$.

Each variable $x_i$ is coupled to its own thermal bath characterized by friction constants $\gamma^{(i)} > 0$ 
and white noises $\zeta_i$, with $\zeta_i = (2k_B T^{(i)} \gamma^{(i)})^{1/2} \xi_i$.
Here $\xi_i$ is a Gaussian-distributed noise with zero mean and correlation
$\langle \xi_i(t) \xi_j(t') \rangle = \delta_{ij} \delta(t - t')$, where $\delta_{ij}$ denotes the Kronecker symbol, and $\langle \ldots\rangle $ denotes an average over noise realizations.
Further, $k_B$ denotes the Boltzmann constant, and $T^{(i)} $ represents the temperature of the bath associated with the component $i$.
For later use, we note that $\gamma^{(i)}$ and $T^{(i)}$ may be seen as principal components of the diagonal
friction tensor $\bm{\gamma}$ and temperature tensor $\mathbf{T}$, respectively (with elements $\gamma_{ij}=\gamma^{(i)}\delta_{ij}$ and $T_{ij}=T^{(i)}\delta_{ij}$).
%

Assuming overdamped motion, the dynamics of the components of $\mathbf{x}$ is governed by the Langevin Equations (LE) \cite{risken1996fokker}
%
\begin{equation} \label{eq:langequ}
\dot{x}_i = (\gamma^{(i)})^{-1} \left(F_i + \zeta_i\right),
\end{equation}
which, thanks to the linearity of the forces, corresponds to a generalized Ornstein-Uhlenbeck process \cite{risken1996fokker}.

For convenience, we henceforth work with a dimensionless LE. To this end we first scale all force coefficients, as well as friction constants and temperatures,
with appropriate reference values. This means that $k_B$ disappears. We then absorb the dimensionless friction coefficients $\gamma^{(i)}$ into the 
dimensionless coefficients $a^{(i)}$, $M_{i j}$ of the forces $F_i$, and into the  temperatures $T^{(i)}$. Specifically, we set
\begin{eqnarray}
\label{eq:rescaling}
\left(\gamma^{(i)}\right)^{-1}F_i & \rightarrow & F_i,\nonumber\\
\left(\gamma^{(i)}\right)^{-1}T^{(i)} &\rightarrow & T^{(i)}.
\end{eqnarray}
With this, the rescaled Langevin equations read
  \begin{equation} \label{eq:langequzw}   
\dot{x}_i  = F_i+\zeta_i= - a^{(i)} x_i + M_{i j } x_j + \zeta_i=K_{ij}x_j+\zeta_i,
 \end{equation} 
 where $\zeta_i  = ( 2 \, T ^{(i)} )^{1/2} \xi_i$, and $\langle   \xi_i(t)  \,  \xi_j(t')\rangle  = \delta_{i j} \, \delta(t - t') $. In compact form, Eq.~(\ref{eq:langequzw})
 can be written as
\begin{align}
    \dot{\mathbf{x}}= -\mathsf{a}\,\mathbf{x}+\mathbf{M} \, \mathbf{x}+\bm{\zeta}=\mathbf{K}\, \mathbf{x}+{\bm{\zeta}}.
 \label{eq:Langevin}
\end{align}

For $\mathbf{M}= 0$ and $a^{(i)}>0$, the system approaches equilibrium, and it is easily shown that the long-time behavior ($t\to\infty$) of the equal-time correlations function $\langle x_i (t)x_j (t)\rangle $ is given by
\begin{align} 
\label{eq:avequ}
\lim_{t \to \infty } \langle x_i(t)x_j(t) \rangle =\langle x_i \, x_j\rangle_{\text{eq}} \equiv \delta_{ij}\frac{T^{(i)} }{a^{(i)}}.
\end{align}
In contrast, when $\mathbf{M} \neq 0$, the averages of the bilinear quantities $x_i^2 $ and $x_i \, x_j$ deviate from their equilibrium values, as demonstrated in the subsequent analysis.

For the choice $n = 3$, on which we later focus,  the matrix $\mathbf{M}$ contains $6$ model parameters. Particular coupling schemes are discussed in 
Refs.~\cite{Doerries_2021} and \cite{AAPPAAPP.1011A3}. Our choice describing a completely nonreciprocal interaction is specified in Eqs.~(\ref{MCouette}) and (\ref{defbirot}).

To close our introduction of the model, we note that, as an alternative to the set of fully Markovian Langevin equations given in (\ref{eq:langequzw}), one 
may also consider a coarse-grained representation involving only the physically observable variables (in our case, $x_1$ and $x_2$), while the auxiliary variables (here: $x_3$) are "integrated out".
This leads to a non-Markovian representation of the dynamics that is explicitly derived for the case $n=3$ in Appendix~\ref{sec:projection}. There we obtain the generalized Langevin equation (\ref{eq:gle})
involving an exponentially decaying memory kernel
and colored noise. Importantly, in the present system, these two functions are not automatically linked via a fluctuation-dissipation theorem [see Eq.~(\ref{eq:FDT3})], consistent with other models of non-Markovian systems under shear \cite{PhysRevE.107.064102}.
\subsection{\label{sec:fokker}Fokker-Planck equation}
In the present paper we are mainly interested in studying the dynamics on the ensemble level, that is, via averages over many realizations of the noise.
We thus consider the time evolution of the normalized probability density  $\rho(\mathbf{x},t)$  with $\int_V \rho(\mathbf{x}, t) d\mathbf{x} = 1$, $t \geq 0$, 
where $d\mathbf{x}$ is the $n$-dimensional integration element. As a conserved quantity, $\rho(\mathbf{x},t)$ fulfills the continuity equation
\begin{equation} 
\label{eq:conteq}
\partial_t \rho(\mathbf{x},t) + \nabla \cdot \mathbf{j}(\mathbf{x},t) = 0,
\end{equation}
where the nabla operator $\mathbf{\nabla}$ has components $\nabla_i = \partial /\partial x_i$, $\nabla\cdot$ denotes the divergence, and $\mathbf{j}(\mathbf{x},t)$ is the current density. 
In the case of diffusion in the presence of a force $\mathbf{F}(\mathbf{x},t)$
%
one has (before rescaling) \cite{risken1996fokker,doi1988theory}
\begin{equation} \label{eq:j}
\mathbf{j}= \mathbf{j}(\mathbf{x},t) = {\bm{\gamma}}^{-1}\left(\mathbf{F}(\mathbf{x},t) \rho(\mathbf{x},t) - k_B \mathbf{T} \nabla \rho(\mathbf{x},t)\right),
\end{equation}
where ${\bm{\gamma}}^{-1} k_B \mathbf{T}=\mathbf{D}$  is the (diagonal) diffusion tensor. Inserting Eq.~(\ref{eq:j}) into the continuity equation (\ref{eq:conteq}) yields the Fokker-Planck (FP) or Smoluchowski equation 
\cite{risken1996fokker,doi1988theory}. In analogy to the LE~(\ref{eq:langequzw}) we use dimensionless variables (thus, $k_B$ disappears), and
absorb the friction coefficients $\gamma^{(i)}$ into forces and temperatures according to (\ref{eq:rescaling}). 
The rescaled FP equation
in component notation reads
\begin{equation}
\label{eq:Smolscalzw}
\partial_t  \rho +\nabla_i \underbrace{\left( F_i \rho - T^{(i)}  \nabla_i \rho \right)}_{j_i} = 0,
\end{equation} 
where, according to Eq.~(\ref{eq:Fi}),
 $F_i=- a^{(i)} x_i + M_{i k} x_k=K_{ik}x_k$. The currents densities $j_i$ can be rewritten as $j_i(\mathbf{x},t)=\rho(\mathbf{x},t)v_i$ with the mean velocities
%
%
\begin{equation}
\label{eq:velocity}
v_i =    F_i -   T ^{(i)} \nabla_i \ln  \rho,
\end{equation}
where the last term may be considered as an effective (dimensionless) force $F_i^{\mathrm{fluct}}$ representing the impact of fluctuations.
%

In stationary states where $\partial_t \rho(\mathbf{x},t) = \nabla \cdot \mathbf{j} = 0$ (i.e., in equilibrium or in a nonequilibrium steady state), the positional distribution $\rho$ 
is Gaussian \cite{risken1996fokker}, that is,
\begin{equation} \label{eq:Gssgeneral}
\rho = Z^{-1} \exp\left[ -\frac{1}{2} (\mathbf{X}^{-1})_{ij}x_ix_j \right],
\end{equation}
with normalization factor $Z = (2 \pi)^{n/2} \sqrt{\text{Det}(\mathbf{X})}$, where $\text{Det}(\ldots)$ is the determinant.
Here, $\mathbf{X}$ is the matrix of second moments with elements $X_{ij} = \langle x_i x_j \rangle$, and 
$\mathbf{X}^{-1}$ denotes the inverse of this matrix. Note that in Eq.~(\ref{eq:Gssgeneral}),
 $\mathbf{X}$ has to be taken in the stationary state, i.e., from the stationary solution of the corresponding relaxation equations (see Sec.~\ref{sec:averages}).
 A stationary solution exists if the deterministic version of the LE~(\ref{eq:Langevin}) is stable, that is, 
 if the matrix $\mathbf{K}$ of coefficients appearing in (\ref{eq:Langevin}) is negative definite. Equivalently, the matrix $\mathbf{a}-\mathbf{M}$ has to be positive definite.
 This yields the stability condition 
 \begin{equation}
 \label{eq:detK}
 -\mathrm{Det}(\mathbf{K})=\mathrm{Det}\left(\mathbf{a}-\mathbf{M}\right) >  0. 
 \end{equation}
 Explicit stability conditions for the case $n=3$ are given in Eqs.~(\ref{eq:detaM}) for general coupling parameters and
 (\ref{eq:d3_explicit}) for a specific parameter choice. For the two-dimensional system without auxiliary variable, see Eq.~(\ref{D2}).
 
 We also note that Eq.~(\ref{eq:Gssgeneral}) is formulated assuming $\langle \mathbf{x} \rangle = 0$ in the stationary state. 
 This is indeed the case in our system as will be later shown.
 If $\langle \mathbf{x} \rangle \neq 0$ (which may be caused by an additional, spatially constant force) the variable $\mathbf{x}$ in (\ref{eq:Gssgeneral}) 
 and in $F^{\mathrm{fluct}}$ of (\ref{eq:viGssrho}) needs to be replaced by $\mathbf{x} - \langle \mathbf{x} \rangle$.

Finally, we note that for the Gaussian distribution (\ref{eq:Gssgeneral}), the velocities defined in (\ref{eq:velocity}) can be calculated explicitly, yielding
%
\begin{equation}
\label{eq:viGssrho}
v_i = F_i + F_i^{\mathrm{fluct}}=K_{ij}x_j+T^{(i)}(\mathbf{X}^{-1})_{i j}x_j.
\end{equation}
\subsection{\label{sec:source}Boundary conditions: Source-sink set-up}
So far, the FP equation~(\ref{eq:Smolscalzw}) has been formulated choosing, as it is quite common, natural boundary conditions.
That is, the density is conserved within an infinite volume and corresponding to this conservation, the currents vanish on the surface at infinite distance from the (trap) center.

In this paragraph we present an extended FP equation whose form is motivated by our actual numerical calculations based on direct numerical solution of the underlying LE via the Brownian Dynamics (BD) method (for details, see Appendix~\ref{sec:technical}). The application of the BD method is straightforward when the parameters are such that the system approaches a stationary state
[see Eq.~(\ref{eq:detK})].

However, the numerical simulations (and their interpretation) become challenging when the stability conditions are violated, a situation that, as we later show, 
can indeed occur when the external flow and the coupling to the auxiliary variable become sufficiently large.
Physically, this instability implies that the particle can escape from the harmonic trap. 
In the BD simulation,
this leads to numerical overflow. To handle such a situation, we have proposed in \cite{AAPPAAPP.1011A3} a control mechanism that involves a self-regulating \emph{source-sink set-up}.
In practice, this means that the particle is ''caught and removed'' when the distance
$\sqrt{x^2+y^2}=r$ from the origin, that coincides with the center of the trap,
reaches the large, yet finite distance $R_c$ from the trap. After one time step, the particle is then "reinserted" at the center, that is, all of its coordinates are set to zero ($\mathbf{x}=0$).
Practically, the squared distance $R_c^2$ is chosen, at least, about hundred times larger than the quantity $\langle x^2+y^2\rangle$ in the equilibrium state (see Appendix~\ref{sec:technical}).

Therefore, one is dealing with a “source-sink” set up, where the "source" is at the center, and the "sink" corresponds to the surface at $R_c$. The two processes (catching and reinserting) are assumed to be balanced on the average. Our numerical calculations show that this control strategy of the stochastic dynamics indeed allows for a quasi-stationary state within the (spherical) volume confined by $R_c$.
Thus, trajectories, density plots and averages can be numerically computed even when the original stability condition (\ref{eq:detK}) is violated. We note that the control mechanism acts only when an instability occurs.

On the ensemble level, the above strategy implies the presence of an additional source term $q(\mathbf{x})=q(x,y)$ in the FP equation (\ref{eq:Smolscalzw}) that now becomes
\begin{equation}\label{eq:Smolsource} 
\partial_t  \rho  +   \nabla_i  j_i =  q,
\end{equation} 
where $j_i$ is given by (\ref{eq:Smolscalzw}).  To ensure that the probability density is conserved in the {\em finite} volume $\tilde{V}\propto R_c^2$, that is,
$\frac{d}{dt}\int_{\tilde{V}} \rho(\mathbf{x},t) d\mathbf{x} = 0$, we have to require that the "loss" via the currents through the surface of $\tilde{V}$, $\partial \tilde{V}$, is balanced by the "gain"
via the source. Using Gauss' theorem, this implies
\begin{equation}\label{eq:rhonorm}
\int_{\partial \tilde{V}} n_i j_i do = \int_{\tilde{V}} q(x,y) d\mathbf{x} \equiv \nu,
\end{equation}
where, on the left side, $n_i$ represents the component of the outward normal vector of the surface, and $do$ is the surface element. 
On the right side of Eq.~(\ref{eq:rhonorm}), $\nu$ represents the rate of transitions from the source to the sink located at the surface. In the following, the source is assumed to be point-like and located at $r=0$, resulting in $q(\mathbf{x}) = \nu\delta(\mathbf{x})$, where $\delta(\mathbf{x})$ denotes the Dirac delta distribution.
\subsection{\label{sec:averages}Time dependence of averages}
In the following we discuss, for a general number of variables, the time change of averages of the type
\begin{equation}
\label{eq:psi}
\langle  \Psi \rangle (t)=  \int_{\tilde{V}} \Psi(\mathbf{x}) \rho(\mathbf{x},t) d\mathbf{x} \equiv \langle  \Psi \rangle,
\end{equation}
based on the FP equation with source term, Eq.~(\ref{eq:Smolsource}).

For notational ease, we henceforth drop the dependency of $\langle  \Psi \rangle$ on $t$.
Since the functions $\Psi(\mathbf{x})$ appearing in the integral do not explicitly depend on time, the dynamics of the averages
is determined by that of $\rho$, i.e., $\frac{d}{dt} \langle\Psi\rangle = \int_{\tilde{V}} \Psi \, \partial_t \rho \, d\mathbf{x}$. 
Replacing $\partial_t\rho$ by  Eq.~(\ref{eq:Smolsource}), integrating by parts, and using $ \int_{\tilde{V}} j_i \, \nabla_i \Psi \, d\mathbf{x} = \langle v_i \nabla_i \Psi \rangle$ we obtain
the relaxation equation
\begin{equation}
\label{eq:dtpsi}
\frac{d}{dt} \langle \Psi \rangle =
\langle v_i \nabla_i \Psi \rangle - \int_{\partial {\tilde{V}}} n_i j_i \Psi do + \int_{\tilde{V}} \Psi q d\mathbf{x}.
\end{equation}
In the special case $\Psi(\mathbf{x}) = 1$, relation~(\ref{eq:dtpsi}) simplifies to Eq.~(\ref{eq:rhonorm}). We also note that
the integral $\int_{\tilde{V}} \Psi  q d\mathbf{x}$ involving the source term vanishes when $q \sim \delta(\mathbf{x})$ (as already assumed  before) and, moreover,
when $\Psi(0) = 0$. This is the case considered from now on.

For spatially varying functions $\Psi(\mathbf{x})$,
the first term on the right-hand side of Eq.~(\ref{eq:dtpsi}) can be rewritten by employing again Eq.~(\ref{eq:Smolsource}) and integrating by parts, yielding 
\begin{eqnarray}
\label{eq:dtpsizw}
\langle v_i \nabla_i \Psi \rangle &=& K_{ik} \langle x_k \nabla_i \Psi \rangle \nonumber\\
& &+ T_{ik}\left(\langle \nabla_k \nabla_i \Psi \rangle - \int_{\partial \tilde{V}} n_k  (\nabla_i \Psi)  \rho do\right).
\end{eqnarray}
We recall that $T_{ij}=T^{(i)}\delta_{ij}$ are the elements of the temperature tensor $\mathbf{T}$ introduced before Eq.~(\ref{eq:langequ}). Thus,
 $T_{ik}\nabla_k=\sum_k T^{(i)}\delta_{ik}\nabla_k=T^{(i)}\nabla_i$.

We now focus on the time evolution (relaxation) of the first and second moments of the distribution function, that is, $\Psi(\mathbf{x}) = x_{\ell}$ and $\Psi(\mathbf{x}) =  x_{\ell} x_j$, respectively.

For the first moment we find from Eq.~(\ref{eq:dtpsi})
\begin{equation}\label{eq:dtxlin}
\frac{d}{dt} \langle x_{\ell} \rangle = \langle v_{\ell} \rangle - \sigma_{\ell} \, , \quad
\sigma_{\ell} = \int_{\partial \tilde{V}} n_i j_i x_{\ell} \, do \, ,
\end{equation}
where $\langle v_{\ell} \rangle$ is defined as
\begin{equation}\label{eq:dtxlinzw}
\langle v_{\ell} \rangle = K_{\ell k} \langle x_k \rangle - T_{\ell k} \int_{\partial\tilde{V}} n_k \rho  do,
\end{equation}
and the elements of the matrix $\mathbf{K}$ are defined in Eq.~(\ref{eq:langequzw}).

%
From Eqs.~(\ref{eq:dtxlin}) and (\ref{eq:dtxlinzw}) it follows that in the absence of boundary terms, the dynamics of the first moment is determined by $\langle v_{\ell} \rangle=K_{\ell k} \langle x_k \rangle $ alone. 
Equation~(\ref{eq:dtpsi}) then implies that, if $\mathbf{K}$ is negative definite (corresponding to the existence of a stationary solution, see Eq.~(\ref{eq:detK})), all first moments $\langle x_{\ell} \rangle$ relax to zero.
Physically, this is expected due to the absence of a constant force in our model.

We thus concentrate on the second moments $\langle x_{\ell} x_j \rangle=X_{\ell j} $ which form the elements of the matrix $\mathbf{X}$ introduced in Eq.~(\ref{eq:Gssgeneral}).
Equation~(\ref{eq:dtpsi}) yields the relaxation equation
\begin{equation}\label{eq:dtxxbil}
\frac{d}{dt} X_{\ell j} = \langle v_{\ell} x_j \rangle + \langle v_{j} x_{\ell} \rangle - \sigma_{\ell j},
\end{equation}
where $\sigma_{\ell j}$ are the elements of the matrix $\bm{\sigma}$ defined as
\begin{equation}
\sigma_{\ell j} = \int_{\partial V} n_i j_i x_{\ell} x_j do \, .
\end{equation}
Note that $\bm{\sigma}$ is symmetric. Further,  $\langle v_{\ell} x_j \rangle$ is defined as
\begin{equation}\label{eq:dtxxbilzw}
\langle v_{\ell} x_j \rangle = K_{\ell k} X_{k j} + T_{\ell j}- \int_{\partial V} n_k T_{\ell k} x_j \rho  do \, .
\end{equation}

We finally note that the relaxation equations (\ref{eq:dtpsi}), (\ref{eq:dtxlin}) and (\ref{eq:dtxxbil}) are {\em exact}; they do not rely on the assumption of Gaussianity of the (steady-state) probability distribution, not even on the linearity of the forces in model. In fact, the linearity enters only in Eqs.~(\ref{eq:dtxlinzw}) and (\ref{eq:dtxxbilzw}).
It is clear, however, that a major complication of all the relaxtion equations lies in the calculation of the boundary terms. These become relevant (only) in the context of the source-sink set-up controlling instabilities.
In fact,
even for linear forces the presence of a source term can induce non-linearity, and thus, the steady-distribution function for this case is generally unknown. Moreover, already without boundary terms,
the actual dependencies of measurable quantities on the the model parameters can be strongly nonlinear, as we will demonstrate in Sec.~\ref{sec:results}.
\section{\label{sec:specific_model} Explicit expressions} 
In this section we apply the expressions derived so far, which were valid for arbitrary (finite) $n$ and arbitrary linear coupling, to the system at hand:
a Brownian particle with two physical (i.e., observable) degrees of freedom ($x_1=x$ and $x_2=y$) that is confined to a harmonic optical trap, externally driven by a volume-conserving shear flow, and is coupled to an auxiliary (hidden) variable $x_3$. 

We focus on the case of {\em one} auxiliary variable for several reasons. Most importantly, this provides us with the simplest mathematical model for a memory effect induced by complex medium, 
namely an exponentially decaying memory function with a single relaxation time (see Appendix~\ref{sec:projection}). Further, the same approach has been used earlier, e.g., in the contexts of reaction kinetics \cite{kappler2018memory} and superdiffusion \cite{siegle2010markovian}, the modeling of active particles with fluctuating self-propulsion \cite{Dabelow2019}, and as a 
minimal model for time-delayed feedback \cite{Loos_2020}. 
Please note that in all of these contexts, the auxiliary variable is not seen as an actual particle, it rather serves as a tool describing, on minimal grounds, the presence of a complex medium.
Adding more auxiliary variables provides no conceptual problem (see \cite{AAPPAAPP.1011A3} for a general discussion). For example, already two auxiliary variables allow to describe more complicated (even oscillatory) memory functions which may be needed to model (or even fit) the behavior of certain correlation functions with several time scales \cite{Doerries_2021,berner2018oscillating}.
On the other hand, when it comes to thermodynamic notions, we do not expect fundamental differences \cite{Loos_2020,loos2021medium} when extending the model by more than one auxiliary variable. We therefore stick to the present model.

After specifying the parameters for this $n=3$-dimensional system in Sec.~\ref{sec:parameter}, we present in Sec.~\ref{sec:solutions} explicit expressions for stability conditions, and for the stationary solutions of the relaxation equations in the absence of boundary terms. Since we are dealing with a fully linear model, the stationary solution is Gaussian, see Eq.~(\ref{eq:Gssgeneral}),
and we can calculate all moments exactly.
\subsection{\label{sec:parameter}Parameters}
We start with the elements of the matrix $\mathbf{M}$ [introduced in Eq.~(\ref{eq:vector_F})]
characterizing the forces beyond the restoring force from the trap. As noted before, $\mathbf{M}$ involves two types of couplings.

First, the physical variables $x_1=x$ and $x_2=y$ are coupled through a divergence-free (i.e., volume-conserving) flow field $\mathbf{v}(x,y)$ with components $v_x(x,y)$ and $v_y(x,y)$ whose (constant) gradients 
determine the coefficients $M_{12}$ and $M_{21}$. 
%
%
We focus on a plane \emph{Couette} flow, where $\mathbf{v}$ points in $x$-direction, while its gradient points in $y$-direction and is characterized
by the constant shear rate $\Gamma$, that is,
\begin{eqnarray}\label{MCouette}
M_{12} & = & \frac{\partial v_x}{\partial y}=\Gamma,\nonumber\\
 M_{21} & = &\frac{\partial v_y}{\partial x}=0.
\end{eqnarray}
Equations~(\ref{MCouette}) show that $M_{12} \neq M_{21}$, a typical feature of Couette flow (note that symmetric coupling, i.e., $M_{12}=M_{21}$, could be realized as well, using a squeeze (extensional) flow geometry).
The Couette flow represents the first type of nonreciprocal coupling considered in this work. We focus on positive shear rates, $\Gamma>0$.

The second type of force involved in $\mathbf{M}$ is the coupling between $x_1=x$, $x_2=y$, and the auxiliary ("hidden") variable $x_3$ in the three-dimensional system. Here we specialize to a 
rotation-like, antisymmetric (and, thus, nonreciprocal) coupling with angular velocities $\Omega_1$ and $\Omega_2$, that is,
\begin{eqnarray}\label{defbirot}
 M_{23} &= &\Omega_1 = - M_{32}, \nonumber\\ 
 M_{31} &=&    \Omega_2 = - M_{13} .
   \end{eqnarray} 
For other types of couplings, see \cite{AAPPAAPP.1011A3}. Our motivation for the choice (\ref{defbirot}) is the following: as an inspection of the stationary solutions of the relaxation equations (see Sec.~\ref{sec:solutions}) reveals, 
the coupling parameter(s) can be chosen such that, at zero shear, $\langle x y\rangle  = 0$  and, furthermore, $\langle x^2\rangle  - \langle y^2\rangle  = 0$
[see Eq.~(\ref{eq:avxyCoubirotzw}) and (\ref{eq:avx2my2Coubirot})],
just as in an equilibrium system. With this choice, the effect of the hidden variable $x_3$ becomes apparent only when the shear flow is turned on. 


Regarding the restoring forces, we typically assume that all spring constants are equal, i.e., $a^{(i)}=a>0$, $i=1,2,3$.
Further, if not stated otherwise, the temperatures are assumed to be equal as well, such that $T^{(i)}=T$. A typical choice in our actual calculations (see Sec.~\ref{sec:results}) is $a=T=1$.

%
\subsection{\label{sec:solutions}Solutions for the stationary state}
The general condition for stability (and thus, existence of a stationary state) is given by Eq.~(\ref{eq:detK}). The expression for the determinant $D_3=\mathrm{Det}(\mathbf{a}-\mathbf{M})$ in the three-dimensional system with arbitrary coupling parameters is given in Eq.~(\ref{eq:detaM}).
For the parameters introduced in Sec.~\ref{sec:parameter}, we obtain 
\begin{equation}
\label{eq:d3_explicit}
D_3=a^3  + a \Omega_1^2+a\Omega_2^2  -  \Gamma   \Omega_1\Omega_2.
\end{equation}
Inspection of Eq.~(\ref{eq:d3_explicit}) reveals that, for fixed $a$, and fixed $\Omega_1$, $\Omega_2$, $D_3$ becomes zero at the critical shear rate
    \begin{equation}\label{eq:gamcrit} 
  a^{-1} \Gamma_{\mathrm{crit}}=   \frac{\Omega_1^2+\Omega_2^2}{\Omega_1\Omega_2} + \frac{a^2}{\Omega_1\Omega_2} . 
      \end{equation}  
 In contrast, for $\Gamma<\Gamma^{\mathrm{crit}}$, the determinant is positive. This defines the stable regime that allows for a stationary solution of the Fokker-Planck equation.
 We note that there is no critical shear rate in the absence of the coupling of  - both or one of - the physical variables $x_1$, $x_2$ to the auxiliary variable $x_3$
 [see Eq.~(\ref{D2})]. This is indirectly also reflected by Eq.~(\ref{eq:gamcrit}) which predicts a divergence of $\Gamma_{\mathrm{crit}}$
 when one (or both) of the coupling constants $\Omega_i$, $i=1,2$,
 approach zero.
For simplicity, we often set $\Omega_1=\Omega_2=\Omega$. For $a = 1$ and $\Omega = 1$, we then find $\Gamma_{\mathrm{crit}} = 3$. 

Within the stationary regime, the first moments decay to zero [as discussed below Eqs.~(\ref{eq:dtxlin}) and (\ref{eq:dynlinav})]. 
Exact expressions for the steady-state values of bilinear averages can be found by setting to zero the time derivatives of the corresponding relaxation equations given in
Appendix~\ref{ssec:bilav} (including the case of a two-dimensional system without auxiliary variable).

Here we focus on averages involving the physical variables, $x$ and $y$. Plugging in the parameters given in Sec.~\ref{sec:parameter} and setting $\Omega_1=\Omega_2=\Omega$, one obtains
 \begin{align}
  R^2 & =    \langle x^2\rangle   + \langle y^2\rangle    =  \frac{ H^{(0)}  }{{\cal D}} \frac{T}{a},  \label{eq:rad2Coubirotzw} \\
 Q_+/2 & =     \langle x y\rangle=\frac{\Gamma}{a}    \frac{ H^{(1)}  }{{\cal D}}  \frac{T}{a},  \label{eq:avxyCoubirotzw} \\
Q_- & =    \langle x^2\rangle   -  \langle y^2\rangle = \frac{\Gamma^2}{a^2}  \frac{ H^{(2)}  }{{\cal D}}  \frac{T}{a},   \label{eq:avx2my2Coubirot} 
   \end{align}    
 where
 \begin{align}
   H^{(0)} &=  a^{-5} ( 16  a^5 + 40   a^3  \Omega^2 + 16 a  \Omega^4  -   \Gamma (10 a^2 \Omega^2 + 2 \Omega^4)  \nonumber\\
 &+ \Gamma^2 ( 4 a^3   + 5 a  \Omega^2)    - \Gamma^3  \Omega^2    ),  \label{eq:radH0} \\
  H^{(1)} &= a^{-4}  \left(4 a^4 + 8 a^2   \Omega^2 + 3  \Omega^4 - a \Gamma  \Omega^2 \right),\\
 H^{(2)} &= a^{-3} (4 a^3 + 5 a  \,  \Omega^2 - \Gamma  \Omega^2 ),
    \end{align} 
and the quantity appearing in the denominator of Eqs.~(\ref{eq:rad2Coubirotzw})-(\ref{eq:avx2my2Coubirot}) is given by
  \begin{equation}
  \label{eq:avxyCoubirotDet} 
 {\cal D} =   a^{- 6} \left(a^3  + 2 a \Omega^2  -  \Gamma   \Omega^2\right) \left( 8 a^3 + 4 a  \Omega^2  +  \Gamma   \Omega^2\right).
      \end{equation}  
We note that  ${\cal D}$ becomes zero at the same critical shear rate $\Gamma^{\mathrm{crit}}$ given in Eq.~(\ref{eq:gamcrit}), as $D_3$.

Physically, $R^2$ corresponds to the squared "radius of gyration" measuring the spatial extent of the probability cloud, while $Q_{\pm}$ characterize the quadrupolar deformation 
of the probability cloud (note that there is no dipolar deformation due to the vanishing of the first moments). We also see that $Q_{\pm}$ vanishes in the limit $\Gamma\to 0$ even in presence of the auxiliary variable ($\Omega\neq 0$). This conforms with the expectation for these quantities in an equilibrium system.
\section{\label{sec:thermo} nonequilibrium thermodynamics and relation to mechanical properties}
In this section we introduce, first, mechanical properties of the system (Sec.~\ref{sec:mechanics}). We then move in Sec.~\ref{sec:entropy} towards quantities from nonequilibrium thermodynamics which, as we will see, are closely related.
\subsection{\label{sec:mechanics} Mechanical properties}
We here consider the torque and angular momentum, which are known to provide interesting information in nonequilibrium systems. An example is a Brownian gyrator \cite{filliger2007brownian,cerasoli2018asymmetry},
that is, a particle coupled to two heat baths, where temperature differences can induce a spontaneous average torque \cite{filliger2007brownian}. Another example are elongated objects, such as polymers, 
under shear flow \cite{aust2002rotation}.

Considering first the $n$-dimensional case, generalized angular momenta and torques may be defined in terms of second-rank tensors $\mathbf{L}$ and $\mathbf{N}$ with elements
\begin{eqnarray}
L_{ij} &=&      \langle  x_i v_j  \rangle   -  \langle  x_j v_i  \rangle,\label{def:NL}\\
N_{ij} & = &      \langle  x_i \gamma^{(j)} F_j  \rangle   -  \langle  x_j  \gamma^{(i)}  F_i  \label{def:NL2}\rangle.
\end{eqnarray}
%
Both quantities have been defined in analogy to corresponding expressions (involving cross products) from classical mechanics. 
The factor $\gamma^{(i)}$ in Eq.~(\ref{def:NL2}) has been inserted to compensate for our rescaling of the forces, see Eq.~(\ref{eq:rescaling}). 
The velocities entering the angular momentum are given by Eq.~(\ref{eq:velocity}), which includes the fluctuating force
$F^{\mathrm{fluct}}_i\propto \nabla\ln\rho$. Alternatively, one may use the LE itself, setting $v_i=\dot{x}_i$ as given in Eq.~(\ref{eq:langequzw}). This is the route in our numerical calculations.
In contrast to the angular momentum, the torque involves only the deterministic part of the force, $F_i$.
Note that, since we are working in the overdamped limit, velocities and forces are proportional to one another, therefore the torque is {\em not}
the time derivative of the angular momentum. 

By definition,  $\mathbf{L}$ and $\mathbf{N}$ are antisymmetric. For the case $n=3$, the matrix elements $\mathbf{L}$ and $\mathbf{N}$  are directly related to one of the three components of the usual angular momentum and torque {\em vectors}.
For example, the third component of the angular momentum vector $\mathbf{\ell}$ is given as $\ell_3 = L_{1 2}$, the other components follow by cyclic permutations.
 For $n=2$, the only non-vanishing component of $\mathbf{L}$ is $L_{12}={\ell}_3$ (and $L_{21}=-{\ell}_3$).

For later use, we consider once again the product $\langle x_k v_i \rangle $ appearing in Eq.~(\ref{def:NL}).
It is useful to decompose this quantity
into anti-symmetric and symmetric parts. Indeed, any second-rank tensor $\mathbf{C}$ can be decomposed according to
$\mathbf{C}=\mathbf{C}^a+\mathbf{C}^s$, where $\mathbf{C}^a$ with
 elements $C_{i k}^a = (1/2) (C_{i k} - C_{k i})$ and  $\mathbf{C}^s$ with elements $C_{i k}^s = (1/2) (C_{i k} + C_{k i})$
is the anti-symmetric and symmetric part, respectively. It is easily seen that $\langle x_k v_i \rangle^a$ is 
proportional to the angular momentum, Eq.~(\ref{def:NL}). Further, $\langle x_k v_i \rangle^s$ appears in the relaxation equation for the second moments, Eq.~(\ref{eq:dtxxbil}).
This yields
\begin{eqnarray}
\label{eq:decompxv}
 \langle x_k  v_i \rangle &=& \langle x_k v_i \rangle^a + \langle x_k v_i \rangle^s\nonumber\\
 & =&\frac{1}{2}L_{k i} +\frac{1}{2}\left(  \frac{d}{dt}  X_{k i} + \sigma_{k i}\right )
 \end{eqnarray}  
involving the surface contribution $\bm{\sigma}$ [see Eq.~(\ref{eq:dtxxbil})].

Equations~(\ref{def:NL}), (\ref{def:NL2}), and the first line of Eq.~(\ref{eq:decompxv}) are general in the sense that they hold for nonlinear deterministic forces $F_j$ as well. Expressions for the linear system at hand, with arbitrary coupling parameters,
are given in Eqs.~(\ref{eq:LX}) and (\ref{eq:NX}) in Appendix~\ref{sec:angular_general}. Due to the linearity, both $\mathbf{L}$ and $\mathbf{N}$ are related to the elements of the matrix of second moments, $\mathbf{X}$.

Here we focus on two coupling scenarios, that will also be discussed later in the results section~\ref{sec:results}. 
We first consider the case $n=2$ without boundaries and in the steady state. Using Eq.~(\ref{eq:LX}) with the explicit expressions for the corresponding second-order moments (see Eqs.~(\ref{mom1122}) and (\ref{mom12}) in the Appendix), and setting $a=(a^{(1)}+a^{(2)})/2$,
$T=(T^{(1)}+T^{(2)})/2$, we obtain for the only non-vanishing component of the angular momentum [see also Eq.~(\ref{eq:LX12dr})], 
\begin{eqnarray}
\label{eq:LX12vi}      
a L_{12}  & = &M_{21} T^{(1)}   -  M_{12}  T^{(2)}=T\left(M_{21}   -  M_{12}\right)\nonumber\\
& &+  \left( M_{21}   +  M_{12} \right) \left( T^{(1)}   -  T^{(2)}\right) .  
\end{eqnarray} 
Equation~(\ref{eq:LX12vi}) directly shows that there are two possible sources of a finite value of the angular momentum and thus, torque.
The first possibility is asymmetry, and thus, non-reciprocity in the mutual coupling between $x_1$ and $x_2$. Secondly, 
even if the coupling is symmetric, different temperatures $T^{(1)}   \neq  T^{(2)}$ can also induce a non-zero torque, in accord with work on Brownian gyrators
\cite{filliger2007brownian,cerasoli2018asymmetry}.
For the special case of linear (Couette) shear flow, where $M_{12}=\Gamma$, $M_{21}=0$, Eq.~(\ref{eq:LX12vi}) reduces to 
 \begin{equation}\label{eq:LCou}      
L_{12}   =    -  \Gamma   T^{(2)}/a.
\end{equation}
Notice the negative sign of $L_{12}$, as well as of the associated angular velocity $w_3=l_{12}/\left(X_{11}+X_{22}\right)$ [with $X_{11}+X_{22}$ playing the role of a moment of inertia].
The angular velocity is a measurable quantity (see, e.g., \cite{doi:10.1021/acs.jpcb.3c02324})
and, as such, particularly relevant. 
These negative signs of $L_{12}$ and ${\ell}_3$ reflect the  
 clock-wise rotational motion induced by the flow, for the  geometry chosen.   

For the full, three-dimensional system involving the auxiliary variable, one finds from Eq.~(\ref{eq:LX}),    
\begin{eqnarray}
\label{eq:LX12}      
{\ell}_3&=&L_{12}   =  -  ( a^{(2)} -  a^{(1)}) X_{12}  \nonumber\\
& & +  M_{21} X_{ 11}  - M_{12} X_{ 22}   +  M_{23} X_{ 31}  - M_{13} X_{ 32}  \,   .  
\end{eqnarray}
The expressions for $L_{23}$ and $L_{31}$ follow from (\ref{eq:LX12}) by cyclic permutation of $1,2,3$. 
%
%
%
For the special case of uniform spring constants ($a^{(i)}=a$), plane Couette flow ($M_{12}=\Gamma$, $M_{21}=0$)
and coupling to the auxiliary variable such that $\Omega_1=\Omega_2=\Omega$, the  components of the angular momentum  are given by 
\begin{eqnarray}
\label{eq:ell123}
\ell_3 &=&  \Omega X_{13}    -   \Gamma X_{22}     +  \Omega X_{23},\nonumber\\
\ell_1 &=  &-  \Omega  X_{22}    +   \Omega  X_{12}  -     \Omega  X_{33},\nonumber\\
\ell_2 &=  &-  \Omega  X_{33}     + \Gamma  X_{23}   -     \Omega  X_{11}  
 +  \Omega   X_{12} .
\end{eqnarray}
Notice that, different to the case $n=2$, the limit of zero shear ($\Gamma\to 0$) does {\em not} imply $\ell_i\to 0$. As we will show below, this (mechanical) effect of the hidden variable has important consequences for the nonequilibrium thermodynamics.
\subsection{\label{sec:entropy} Entropy production rate}
We now turn to a key quantity to measure the system's deviation from equilibrium, that is the rate of change of the total entropy (total entropy production).
This quantity has been considered in various contexts and communities, from macroscopic irreversible thermodynamics \cite{cite-key} to the nonequilibrium dynamics of soft matter systems (such as shear-driven polymers and liquid crystals \cite{doi1988theory,j2007statistical,Hess+1975+728+738}) to
studies in the realm of stochastic thermodynamics, where thermodynamic notions become trajectory-dependent \cite{sekimoto2010stochastic,seifert2012stochastic}. While the general concepts are well established, the present system involves some subtleties
due to the (potentially) different temperatures, nonreciprocity, and boundary effects. We thus repeat some basic notions and then specialize to the system at hand.

We start from the ensemble-averaged total entropy production, $\dot{\Theta}$, that we write as
   %
   \begin{eqnarray}
   \label{entroprod} 
\dot{\Theta}  & = &    
   \int d\mathbf{x}\, \frac{ j_i(\mathbf{x}, t)  j_i(\mathbf{x}, t)}{ T^{(i)}\rho(\mathbf{x},t)}   \nonumber\\
   &=&     \int d\mathbf{x} \,  \frac{ j_i v_i}{T^{(i)}} = 
   \langle   v_i \, \left(T^{(i)}\right)^{-1}   \,   v_i  \rangle \,  ,
\end{eqnarray}
where we have used the relation $j_i=\rho v_i$ (see Eq.~(\ref{eq:Smolscalzw}) below) in the second line. 
In the last member of Eq.~(\ref{entroprod}), is understood that  $ v_i \, \left(T^{(i)}\right)^{-1}   \,   v_i $ stands for $\Sigma_{ i = 1}^n \,  v_i \, \left(T^{(i)}\right)^{-1}   \,   v_i $. Also note that, in numerical calculations, $v_i$ has to be replaced by $\dot x_i(t)$.

Equation~(\ref{entroprod}) directly shows that $ \dot{\Theta}  > 0$ holds true, as it should be (on the average) according to the second law. 
In the framework of stochastic thermodynamics, Eq.~(\ref{entroprod}) may be derived by starting from the time derivative of the stochastic system entropy $s_\text{sys}=-\ln\rho(\mathbf{x},t)$ (see \cite{seifert2005entropy} 
and \cite{Loos_2020} for a system with different heat baths), or by starting from the fluctuating total entropy expressed via path probabilities (see, e.g., \cite{Crisanti2012}).
For uniform temperatures, and back-scaling to reintroduce the friction constant, the expression above becomes consistent with that given in our earlier work \cite{AAPPAAPP.1011A3} 
that conforms, in turn, with \cite{doi1988theory}. 

By using $v_i = F_i - T^{(i)} \nabla_i \ln \rho$ [see Eq.~(\ref{eq:velocity})] for one of the terms in the expression for $v_i$ in Eq.~(\ref{entroprod}), the total entropy production rate can be separated into two contributions associated with the deterministic and the fluctuating forces: 
  \begin{equation}\label{entroprod12} 
 \dot{\Theta} = \dot{\Theta}_1 + \dot{\Theta}_2, 
 \end{equation}
 where
 \begin{eqnarray}
 \label{eq:entropy_parts}
\dot{\Theta}_1  &=&        \langle   F_i \, \left(T^{(i)}\right)^{-1}   \,   v_i  \rangle \label{eq:theta1},\\
 \dot{\Theta}_2 & =&  -  \int d\mathbf{x} \, \left( \nabla_i  \rho \right)   \,   v_i    \label{eq:theta2}.
\end{eqnarray}
Both terms can be further interpreted. In $\dot\Theta_1$, each summand $i$ may be considered as the (averaged) rate of work done by the deterministic force $F_i$ divided by the corresponding temperature.
In the framework of stochastic thermodynamics, using that, for a single trajectory, $F_i=\dot{x}_i-\zeta_i$ [see Eq.~(\ref{eq:langequzw})],
each summand in $\dot\Theta_1$ corresponds, before averaging, to the fluctuating heat exchange $\dot{Q}_i=\left(\dot{x}_i-\zeta_i\right)\dot x_i$
between variable $x_i$ and the surrounding medium due to friction and noise, divided by the corresponding temperature $T^{(i)}$. 
In this sense, $\dot\Theta_1=\langle \dot{Q}_i\rangle/T^{(i)}$ is called the "medium entropy".
In the same framework, $\dot\Theta_2$ represents the ensemble-averaged change of the stochastic system entropy $s_\text{sys}=-\ln\rho(\mathbf{x},t)$, i.e., 
$\dot\Theta_2=\langle \dot{s}_{\text{sys}}\rangle$ (as one may verify following the steps in \cite{Loos_2020} for a multi-temperature system).

%
In a nonequilibrium steady-state, where the probability distribution becomes constant in time, one expects that $\dot\Theta_2=0$. This can be directly verified in the present system where the steady-state distribution
(without boundaries) is Gaussian, see Eq.~(\ref{eq:Gssgeneral}), such that $\nabla_i\rho=\rho\left(\left(-\mathbf{X}^{-1}\right)_{i k} x_k\right)$. Equation~(\ref{eq:theta2}) then becomes
 \begin{eqnarray}
 \label{entroprod2zwgen} 
\dot{\Theta}_2 & =&   \left(\mathbf{X}^{-1}\right)_{i k} \langle x_k \,   v_i   \rangle \nonumber\\
& =& \frac{1}{2} \left((\mathbf{X}^{-1}\right)_{i k} \, L_{k i},
\end{eqnarray}
where we have used Eq.~(\ref{eq:decompxv}). The matrix  $L_{ki}$ is antisymmetric, while $\mathbf{X}^{-1}$ is symmetric (as is $\mathbf{X}$). Therefore, and since we sum over both indices, it follows that $\dot\Theta_2=0$ in the steady state (note, however, that the individual contributions to the sum, which may be related to information flows between subsystems \cite{Loos_2020}, can be nonzero).
%
%
 
 In the steady state (and without boundaries), the total entropy production (\ref{entroprod}) is therefore given by the medium entropy alone, that is,
 \begin{equation}
 \dot\Theta=\dot\Theta_1.
 \end{equation}
 
 We thus consider in more detail Eq.~(\ref{eq:theta1}).
 %
%
 %
In the present model the forces are linear, $ F_i = K_{i k} x_k$, such that $\dot{\Theta_1}$ is determined by the product $\langle x_k v_i \rangle $. 
From Eqs.~(\ref{eq:theta1}) and (\ref{eq:decompxv}) it follows that
\begin{equation}\label{def:dotAzw} 
\dot\Theta_1 =  \frac{1}{2}\left( \frac{1}{T^{(i)}} K_{i k}\right)^a \, L_{k i}   + \frac{1}{2}  \left( \frac{1}{T^{(i)}} K_{i k}\right)^s \,  \left(\frac{d}{dt}  X_{k i} + \sigma_{k i} \right) \,  . 
\end{equation}
 Equation~(\ref{def:dotAzw}) shows the important role of the angular momentum, a purely mechanical property, for the nonequilibrium thermodynamics of the system.
Notice that the contribution associated with $L_{ki}$ is multiplied with the anti-symmetric part of the matrix $(T^{(i)})^{-1} {\mathbf{K}}$, whereas the contributions linked with the time derivative and the surface are multiplied 
with the symmetric part.

In a nonequilibrium steady state, the time derivative $\frac{d}{dt}X$ vanishes. If, furthermore, surface contributions can be neglected, the angular momenta can be easily expressed via the second moments,
see Eq.~(\ref{eq:LX}). We then obtain
\begin{equation}\label{def:dotAsteady} 
\dot\Theta=\dot\Theta_1 =  \frac{1}{2}\left( \frac{1}{T^{(i)}} K_{i k}\right )^a \left( K_{i l} X_{ l k}  - K_{k l} X_{ l i}  \right)
\end{equation}
This result is consistent with expressions for the entropy production rate in Refs.~\cite{Loos_2020,Crisanti2012}.
  \subsubsection{Entropy production rate for $n = 2$ }\label{specn2}  
 Equation~(\ref{def:dotAzw}) simplifies in the absence of the auxiliary variable (i.e., $n=2$). We then have (with $a^{(i)}=a$)
\begin{eqnarray}
\label{def:dotA12}  
\dot\Theta_1 &=&  -  \frac{1}{2} \left( \frac{ M_{12}}{T^{(1)}} -   \frac{M_{21}}{T^{(2)}} \right) \, L_{12} \nonumber\\
& &+  \frac{1}{2} \left( \frac{M_{12}}{T^{(1)}} +  \frac{M_{21}}{T^{(2)}} \right)\,  \left(  \frac{d}{dt}  X_{12} + \sigma_{12} \right).
\end{eqnarray}
 
 In the absence of boundary terms and within the steady state, $\dot\Theta_1$ and, thus, the total entropy production $\dot\Theta$, reduces to the terms given in the first line
 of Eq.~(\ref{def:dotA12}). Replacing $L_{12}$ by Eq.~(\ref{eq:LX12vi}) we find
\begin{eqnarray}
\label{eq:dotA2}
 \dot\Theta=\dot\Theta_1= \frac{1}{2a}\left(M_{12}^2\frac{T^{(2)}}{T^{(1)}}+M_{21}^2\frac{T^{(1)}}{T^{(2)}}-M_{12}M_{21}\right).
 \end{eqnarray}
 In the special case of planar Couette flow ($M_{12}=\Gamma$, $M_{21}=0$) it follows from Eq.~(\ref{eq:dotA2}) 
 \begin{equation}\label{entprod1Cou}  
\dot\Theta= \dot\Theta_1=    \frac{ \Gamma^2}{2a}  \frac{ T^{(2)}}{T^{(1)}} =-\frac{\Gamma}{2aT^{(1)}}L_{12}
\end{equation}
where we have used the result for the angular momentum, Eq.~(\ref{eq:LCou}).
The middle part of Eq.~(\ref{entprod1Cou}) reflects that, for any (positive) temperatures, shear flow (i.e., $\Gamma\neq 0$) leads to a positive entropy production rate, as expected.
The results further depends on the temperature in $x$- and $y$-directions, corresponding to the shear and shear gradient direction. In particular, if the system is "hotter" in the direction of the shear gradient,
the entropy production rate is enhanced compared to the case of uniform temperatures.

%
 
\subsubsection{Entropy production rate for $n = 3$ }\
 For $n = 3$   and when surface contributions do not exist, the expression (\ref{def:dotAzw})  for the entropy production rate becomes (assuming uniform spring constants $a^{(i)}$)
  \begin{equation}\label{dotAell} 
\dot\Theta_1 =- \left(  \frac{M^a_{12}}{T^{(1)}}\,  \ell_3 +   \frac{M^a_{23}}{T^{(2)}}\,  \ell_1 +  \frac{M^a_{31}}{T^{(3)}}\,  \ell_2 \right),
\end{equation}
where ${\ell}_3$ is given in Eq.~(\ref{eq:LX12}), the other follow by cyclic permutation.

We recall that out of the three variables, only two (namely $x_1$ and $x_2$) are considered as physical (observable) variables. The third variable, $x_3$, is considered as an auxiliary (hidden) variable representing the coupling to the medium (inducing memory in a coarse-grained representation, as discussed in Appendix~\ref{sec:projection}).
It is therefore useful to formulate the entropy production rate accordingly, setting $\dot\Theta_1=\dot{\Theta}_1^{\text{obs}}+ \dot{\Theta}_1^{\text{aux}}$.

In the observed system, the only nonvanishing angular momentum is ${\ell}_3=L_{12}$. From Eq.~(\ref{dotAell}) the corresponding contribution to the entropy production rate then follows as
  \begin{equation}\label{dotAell12} 
\dot{\Theta}_1^{\text{obs}}  =-  \frac{M^a_{12}}{T^{(1)}}\,  \ell_3  \,  , 
\end{equation}
The contribution associated with the auxiliary system is 
$ \dot{\Theta}_1^{\text{aux}}  =   \dot\Theta_1- \dot\Theta_1^{\text{obs}}$. 

We now focus on parameters corresponding to planar shear flow and nonreciprocal coupling the auxiliary variable, see Eqs.~(\ref{MCouette}) and Eq.~(\ref{defbirot}).
We further set
$\Omega_1 = \Omega_2 = \Omega$, $ a^{(i)} = a$,  and $T^{(i)} = T$  ($i = 1, 2, 3$). The uniform temperatures are chosen to concentrate on the effect of non-reciprocity of the couplings.
Equation~(\ref{dotAell}) for the total entropy production then becomes (after multiplying with $T$)
 \begin{equation}
 \label{dotAgom}
T \dot\Theta = T\Theta_1= - \frac{1}{2} \Gamma \,  \ell_3 -    \Omega\,  \left(\ell_1 +   \ell_2 \right). 
\end{equation}

Explicit relations between the angular momenta ${\ell}_i$ and the second moments $X_{ij}$ are given in Eqs.~(\ref{eq:ell123}). 
Inserting these expressions and separating the entropy production rate as proposed before, i.e., $\dot\Theta_1=\dot\Theta_1^{\text{obs}} +\dot\Theta_1^{\text{aux}}$,
we obtain
 \begin{eqnarray}
 \label{dotA1}
T\dot{\Theta}_1^{\text{obs}} &=& - \frac{1}{2} \Gamma   \ell_3  \nonumber\\
&=&  \frac{1}{2}\Gamma^2 X_{22} -  
\frac{1}{2} \Gamma  \, \Omega \, \left( X_{13}    +  X_{23}  \right) \, ,   
\end{eqnarray}
and
 \begin{eqnarray}
 \label{dotA2}
 T \dot\Theta_1^{\text{aux}}  &=&  - \Omega \,  \left(\ell_1 +  \ell_2 \right) \nonumber\\
  & =&   -  \Gamma \Omega X_{23}   \nonumber\\
  & & +
\Omega^2  \left(X_{11}   +  X_{22}  +  2  X_{33}  - 2  X_{12}  \right) .  
\end{eqnarray}
Both expressions reveal important insights on their own. For example, $T\dot{\Theta}_1^{\text{obs}}$ reduces to zero for $\Gamma\rightarrow 0$. This is plausible because both of the observable variables 
are directly affected by the shear flow. In contrast,  $T \dot\Theta_1^{\text{aux}} $ can be  non-zero, reflecting that coupling to the hidden variable (and, thus, memory) alone can drive the system away from equilibrium. 
%

%

We close this section with some comments on the actual calculation of the entropy production rate, particularly $\dot\Theta_1$.
As shown before, this quantity can be expressed via the angular momenta, $\ell_i$, that are fully determined by the second moments, see, e.g., Eqs.~(\ref{eq:ell123}). 
 This provides a straightforward way to calculate $\Theta_1$ analytically, at least for shear rates $\Gamma<\Gamma_{\text{crit}}$, where all second moments exist
 (we will later comment on the use of analytical expressions for larger shear rates, see Sec.~\ref{sec:results_entropy}).
 From the numerical side, one may again use (\ref{eq:ell123})  to obtain results for $\dot\Theta_1$ via the computations of the second moments $\mathbf{X}$. This, however, would give no additional insight 
 since for all six second moments $X_{ij}$, the analytical calculation and the numerical results turn out to agree  well for shear rates below $\Gamma_{\text{crit}}$
 (as demonstrated in Figs.~\ref{fig:transR2vsgam} and \ref{fig:transxyqmvsgam}).
An independent numerical calculation of $\dot\Theta_1$ is based on 
relation (\ref{dotAell}) when we calculate the components of the angular momentum directly via their definitions as cross products of position $\mathbf{x}$ and velocity 
$\dot{\mathbf{x}}$, see Eq.~(\ref{def:NL}).
 Finally, for calculation of the total entropy production rate $\dot\Theta$ (which is equivalent to $\dot\Theta_1$ in steady states), we can also evaluate directly the last member of Eq.~(\ref{entroprod}),
 with the replacement $v_i\to \dot x_i(t)$.
\section{\label{sec:results}Results}
In this section we present results from our analytical expressions as well as from numerical calculations based on BD simulations (for details, see Appendix~\ref{sec:technical}).
We concentrate on the case $n=3$ with uniform spring constants and temperatures. Further, we assume planar shear flow (with shear rate $\Gamma$) and uniform coupling with the auxiliary variable characterized by the constant $\Omega$. As discussed in Sec.~\ref{sec:solutions}, the resulting system, in the absence of boundaries, develops a steady state for shear rates in the "precritical" range $\Gamma<\Gamma_{\text{crit}}=3$. 
Beyond $\Gamma_{\text{crit}}$ the deterministic (as well as the stochastic) system is unstable as a consequence of the coupling to the third variable.
However, we can still perform BD simulations using the source-sink set-up described in Sec.~\ref{sec:source}. We recall that this set-up only comes into play for $\Gamma>\Gamma_{\text{crit}}$.
\subsection{\label{sec:results_stationary}Stationary state and instability}
\begin{figure}
	\centering
	\includegraphics[width = 0.48\textwidth]{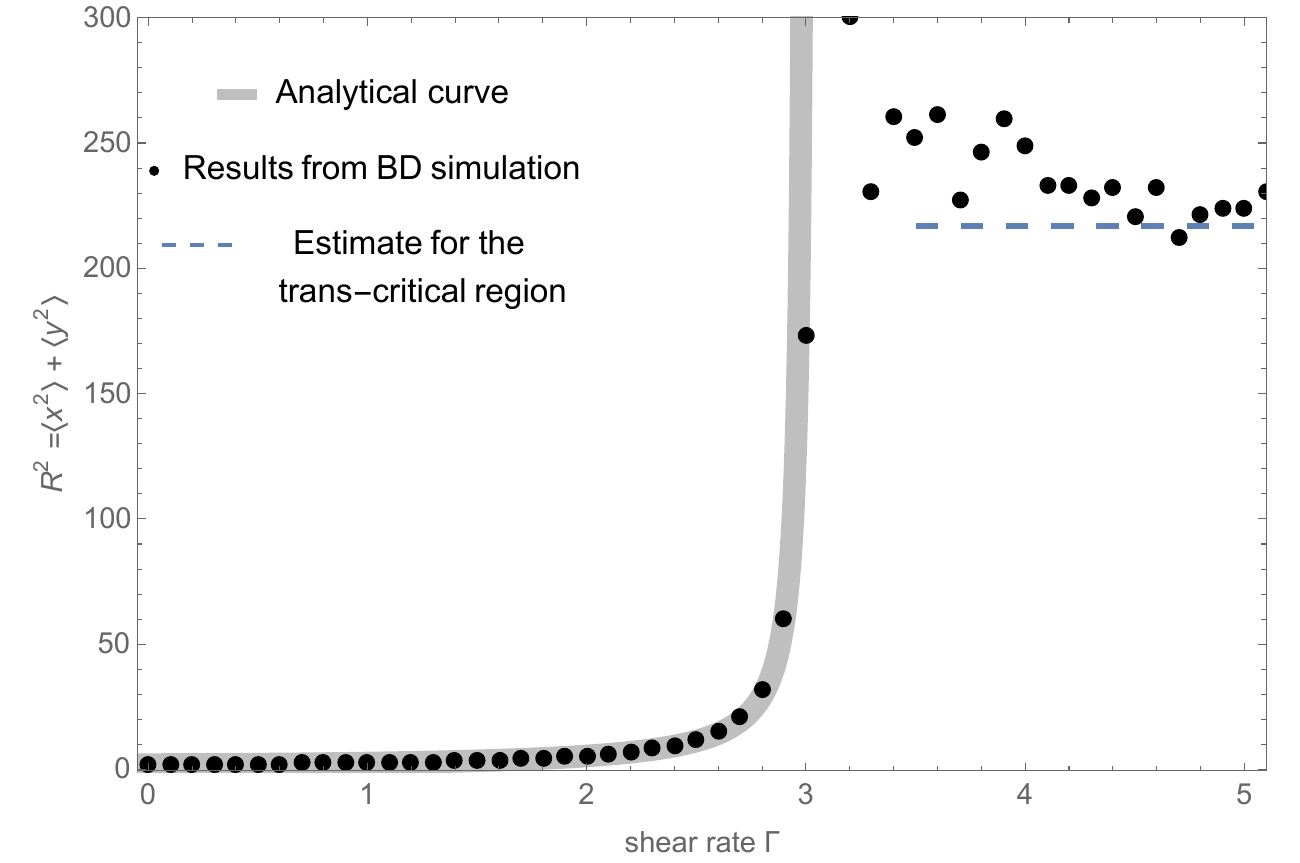}
	\caption{Squared mean radius of gyration, $R^2 = \langle x ^2\rangle  + \langle y^ 2\rangle $, as function of the shear rate. The black points stem from BD simulations, while the thick line follows from the analytical (stationary) solution given in Eq.~(\ref{eq:rad2Coubirotzw}). For the parameters chosen here ($a=T=\Omega=1$), the critical shear rate is $\Gamma _{\text{crit}}= 3.0$.}
	\label{fig:transR2vsgam}
\end{figure}

We start by discussing results for the shear-rate dependency of bilinear averages involving the observable variables $x_1=x$ and $x_2=y$ (note that the corresponding linear averages vanish in our model).
In Fig.~\ref{fig:transR2vsgam} we plot the squared radius of gyration, $R^2 = \langle x^2\rangle + \langle y^2\rangle$, as function of the shear rate. The thick 
lines indicates the analytic result from Eq.~(\ref{eq:rad2Coubirotzw}), while the data points (filled circles) are from BD simulations. Focusing first on the precritical region ($\Gamma<\Gamma_{\text{crit}}$), 
we find that the data are in excellent agreement.
Starting from $\Gamma=0$ and increasing the shear rate, both methods predict a sharp increase of $R^2$ by more than two orders of magnitude. This is a clear indication of the singularity predicted by the analytic solutions of the relaxation equations.
In the transcritical region $\Gamma>\Gamma_{\text{crit}}$, the BD simulations with source-sink set-up still predict {\em finite} (yet somewhat noisy) values for the radius of gyration.
This confirms the performance of our controlled BD simulation method, which is based on the idea that the particle is set back to origin when it reaches the outer rim. The control set-up thus hinders a divergence of the radius of gyration (measuring the spatial extent of the the probability cloud).
In fact, for shear rates far above $\Gamma_{\text{crit}}$, the data appear to saturate at a constant value (independent of the shear rate) which is close 
to the estimate $\tilde{R}^2 \approx R_c^2/ \ln(R_c^2)$ (with $R_c$ being the radius of the rim), as discussed in \cite{AAPPAAPP.1011A3}. We take the saturation of the data as a hint that the controlled system develops some kind of steady state, even if the latter cannot be analytically accessed due to the difficulties in treating the boundary terms.  We also note that
the scatter of the data points is due to the fact that the run time used is not always long enough to yield a sufficiently large number of transits from the center to the rim in order to reach a steady-state behavior. The number of transits is discussed in more detail in \cite{AAPPAAPP.1011A3}.

\begin{figure}
	\centering
	\includegraphics[width = 0.48\textwidth]{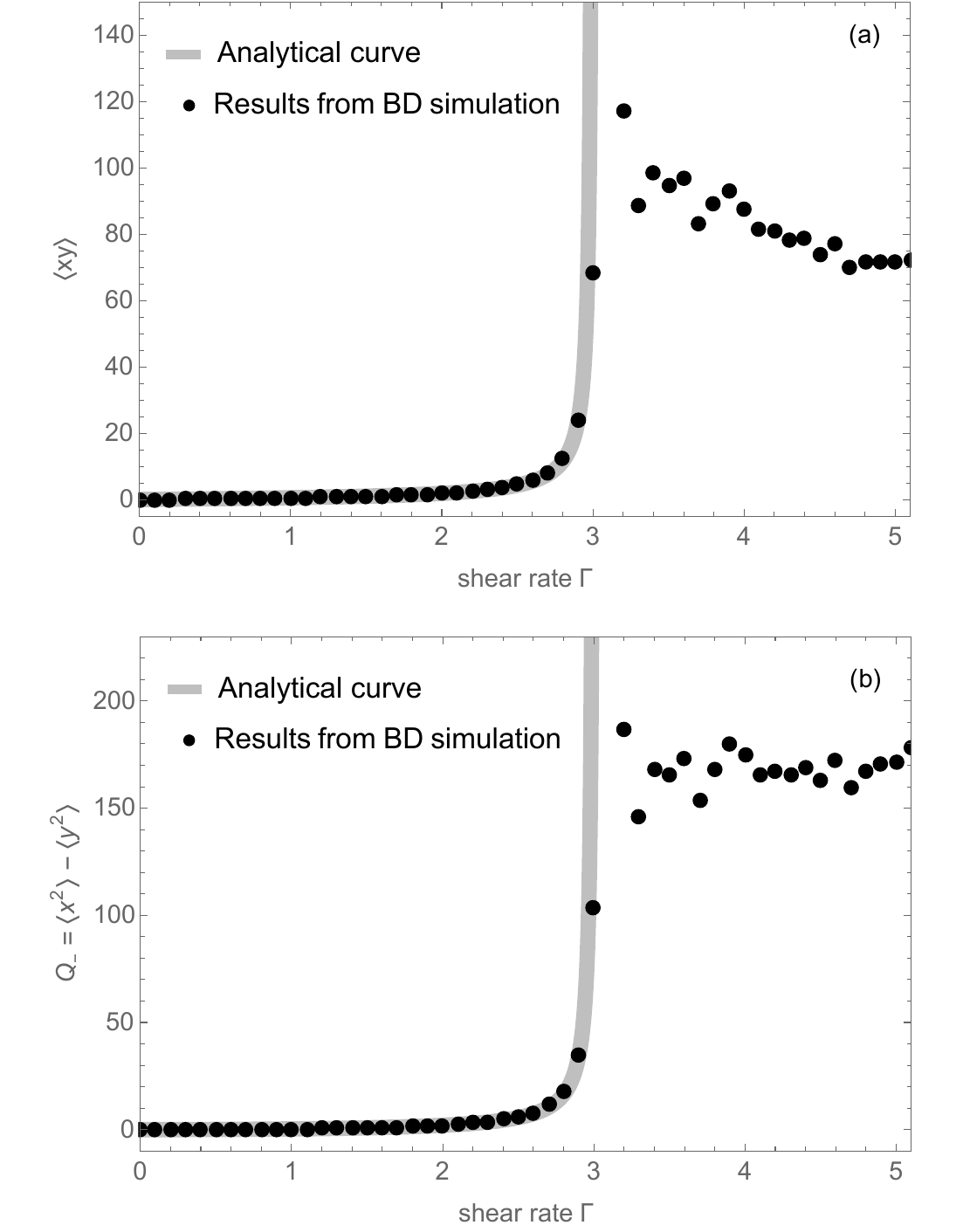}
	\caption{The averages  $\langle x y\rangle $ (a) and $Q_- =  \, \langle x ^2\rangle  -  \langle y^ 2\rangle $ (b)   as function of the shear rate. The black points stem from the BD simulations, while the thick lines follows from the analytical (stationary) solutions given in Eqs.~(\ref{eq:avxyCoubirotzw}) and (\ref{eq:avx2my2Coubirot}). Parameters as in Fig.~\ref{fig:transR2vsgam}.}
	\label{fig:transxyqmvsgam}
\end{figure}

The existence of the singularity is also seen in all the other bilinear averages $X_{ij}=\langle x_i x_j\rangle$. As an illustration, we present in Fig.~\ref{fig:transxyqmvsgam}
results for $\langle x y\rangle$ 
and $Q_- = \langle x^2\rangle - \langle y^2\rangle$. 
Again, we find excellent agreement between the analytical expressions (see Eqs.~(\ref{eq:avxyCoubirotzw}) and (\ref{eq:avx2my2Coubirot}), respectively), and numerical data. Further, the BD data in the transcritical regime confirm the picture of a quasi steady-state stabilized by our control set-up.
\subsection{\label{sec:results_position}Dynamics of positions}
\begin{figure}
	\centering
	\includegraphics[width = 0.55\textwidth]{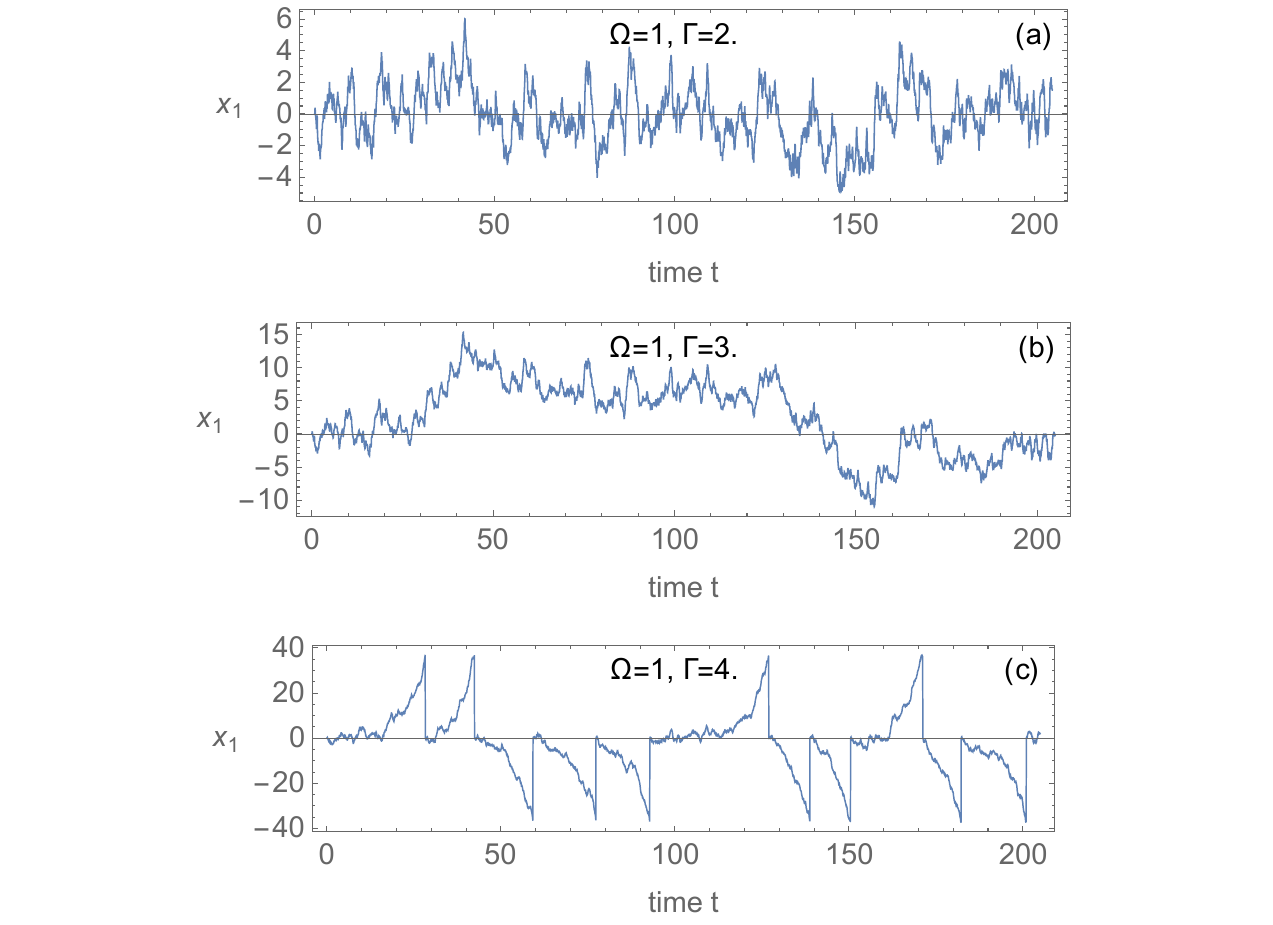} 
\caption{The particle's $x$-position as function of time $t$ at different shear rates: (a) $\Gamma=2$, (b): $\Gamma=\Gamma_{\text{crit}}$, (c): $\Gamma=4$. 
Notice that the scale on the vertical axis increases from (a) to (c). Parameters as in Fig.~\ref{fig:transR2vsgam}. }
\label{fig:xvstom1}
\end{figure}

In view of the dramatic changes of the steady-state averages upon approach of $\Gamma_{\text{crit}}$ (see Figs.~\ref{fig:transR2vsgam} and \ref{fig:transxyqmvsgam}),
it is interesting to study directly the stochastic dynamics of the particle position.
We here focus on the particle's $x$-coordinate as function of time, i.e., $x_1(t)$. As in the preceding section, we consider a system with $\Omega=1$. 
Results for $x_1(t)$ at three different shear rates are shown in Fig.~\ref{fig:xvstom1}. 

In all cases, $x_1$ fluctuates around zero, consistent with the vanishing of the first moment, $\langle x_1\rangle$. However, the character of these fluctuations strongly varies with $\Gamma$.
At low shear ($\Gamma=2$) we observe rapid fluctuations, characterized by short times between sign changes, and relatively small amplitudes. This behavior is similar that observed in a system without coupling to the auxiliary variable
($\Omega=0$, not shown).
Directly at the critical shear rate, $\Gamma_{\text{crit}}=3$, the times between sign changes are much longer indicating a "critical slowing down", while the amplitudes are larger.
Finally, at $\Gamma=4 > \Gamma_{\text{crit}}$, $x_1(t)$ shows a repeated, accelerated increase or decrease with time, followed by a new start at $x_1 =0$. This behavior is in accord with the sink-source set up in our simulations.  
 \subsection{\label{sec:results_secondmoments}Time dependence of second moments}
\begin{figure}
	\centering
\includegraphics[width=0.55\textwidth]{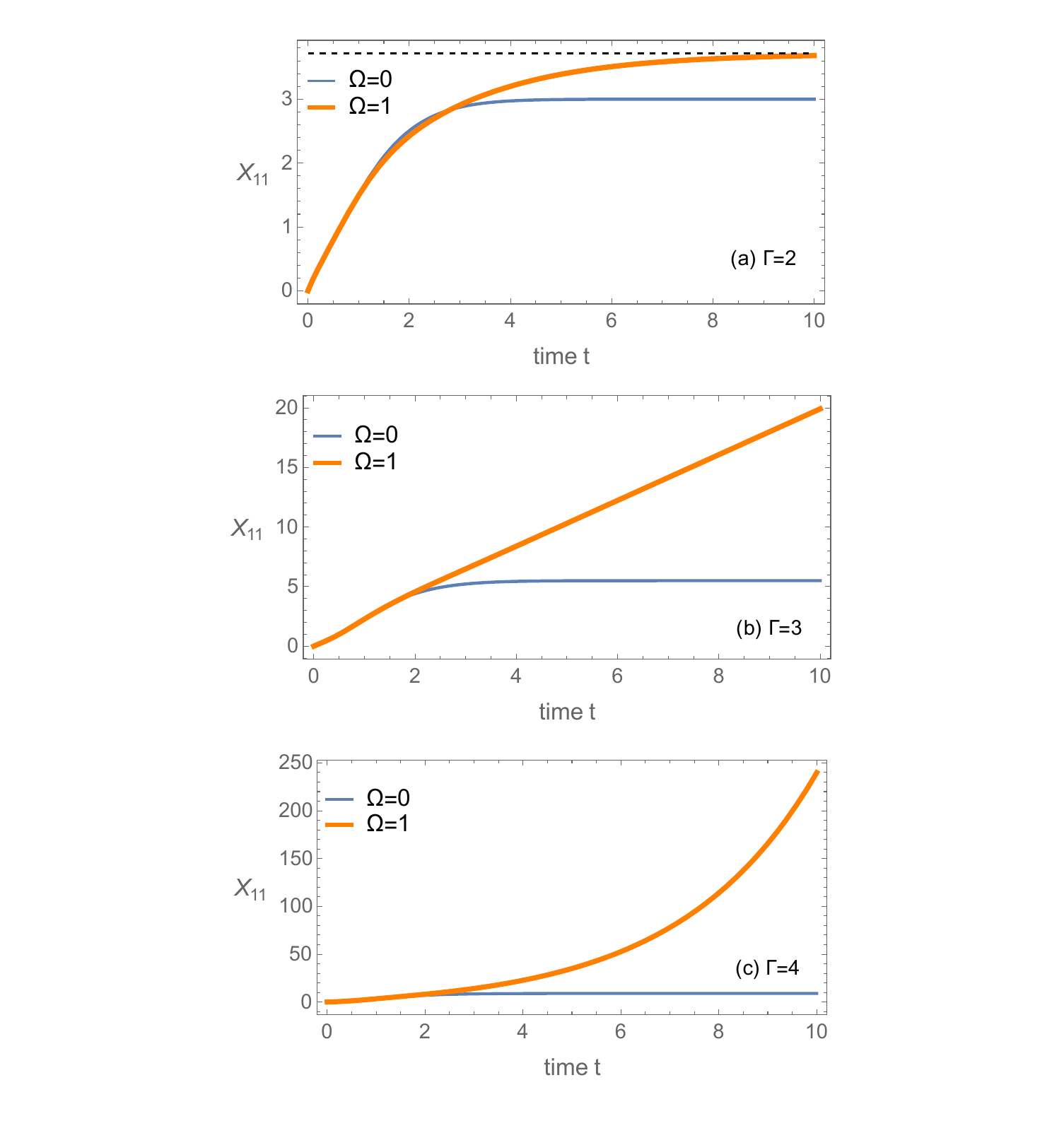} 
\caption{Analytical results for the second moment $X_{11}=\langle x_1^2\rangle$ as function of time for different shear rates. (a) $\Gamma=2$, (b): $\Gamma=\Gamma_{\text{crit}}$, (c): $\Gamma=4$. 
Each panel compares the cases $\Omega=1$ (coupled auxiliary variable) and $\Omega=0$ (no coupling).
The other parameters are as in Fig.~\ref{fig:transR2vsgam}. }
\label{fig:xsqana234}
\end{figure}

A further interesting signature of the dynamics is provided by the time dependence of the second moments. As an example, we here consider the quantity $X_{11}(t)=\langle x_1^2\rangle(t)$.
Analytical results from solution of Eq.~(\ref{eq:bilinrel3D}) are plotted
for three shear rates in Fig.~\ref{fig:xsqana234}, where we have included data for $\Omega=0$. In the latter case, there is no instability and the system reaches a steady state for all values of $\Gamma$. This is confirmed in Fig.~\ref{fig:xsqana234} showing that $X_{11}$ reaches constant final values.

In presence of coupling to the auxiliary variable, a steady state only occurs in the case $\Gamma=2<\Gamma_{\text{crit}}$ (a). Interestingly, the approach of the steady-state value (dashed line)
is significantly slower than in the non-coupled case. For $\Gamma\geq \Gamma_{\text{crit}}$, the effect of the hidden variable becomes very clear.
Directly at the critical shear rate $\Gamma_{\text{crit}}$ (b), the second moment increases essentially linearly in time, resembling "free" diffusive motion. In contrast, for $\Gamma> \Gamma_{\text{crit}}$ one observes an accelerated motion with an exponential increase of the second moment (c).
In the Brownian Dynamics simulations this unbounded increase  is controlled by the source-sink set-up.
Analogous behavior is found for all other second moments $X_{ij}$.
\subsection{\label{sec:results_entropy}Entropy production rate}
\begin{figure}
	\centering
\includegraphics[width=0.48\textwidth]{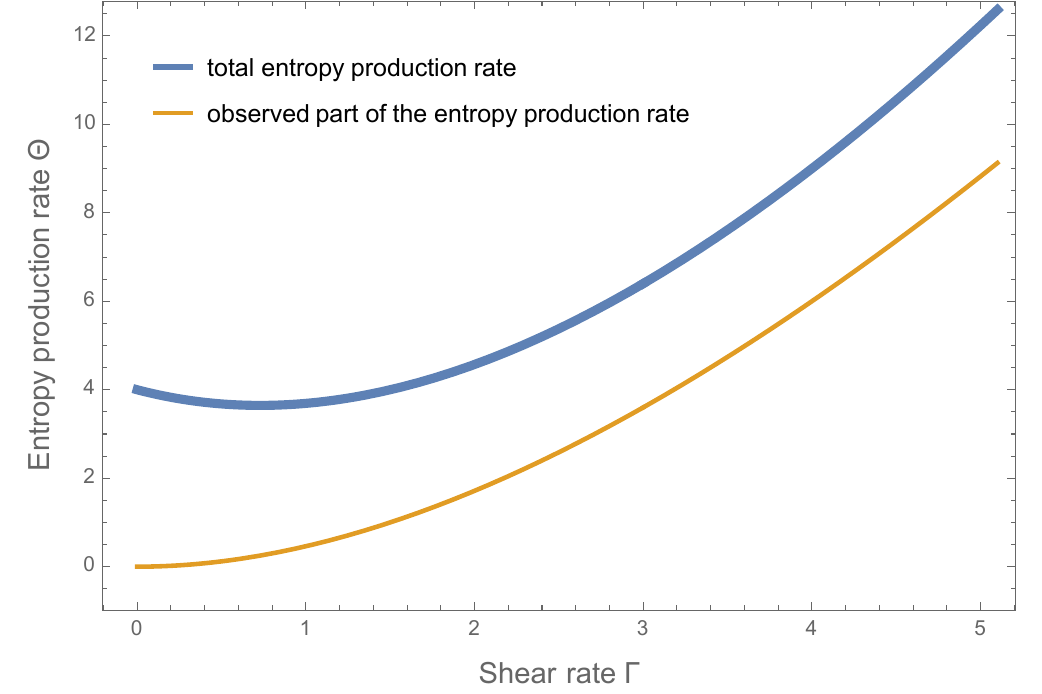}
\caption{ Analytical results for the entropy production rate $\dot{\Theta}=\dot{\Theta}_1$ as function of the shear rate $\Gamma$, based on the stationary solutions of the relaxation equations. 
The  thick blue and orange curves  pertain to the full and the observed part of $\dot{\Theta}$. 
In the precritical range $\Gamma<\Gamma_{\text{crit}}=3$, both quantities can be calculated via their dependency on the second-order moments, or via the angular momenta, see
Eqs.~(\ref{dotAgom}) and (\ref{dotA1}).
The results for $\Gamma\geq \Gamma_{\text{crit}}=3$ follow directly from Eqs.~(\ref{eq:ent_final}) and (\ref{eq:entobs_final}).
Parameters as in Fig.~\ref{fig:transR2vsgam}. }
\label{fig:entrodana}
\end{figure}

In Fig.~\ref{fig:entrodana} we present analytical results for the full and observed entropy production rate, $\dot\Theta$ and $\dot\Theta^\text{obs}$, as functions of $\Gamma$. Since the results are obtained for stationary states, the total entropy production rate equals the medium part, i.e., $\dot\Theta=\dot\Theta_1$.

We start by considering the limit $\Gamma\to 0$. Here, $\dot\Theta$ is non-zero. This reflects that even in the absence of shear, i.e., without external drive, and for uniform temperatures, nonreciprocal couplings alone lead to a finite entropy production, as may already be seen from Eq.~(\ref{dotA2}) for $\Gamma=0$. This observation of "intrinsic nonequilibrium" conforms with other studies \cite{Loos_2020}.
In contrast, the contribution related to the observable variables, $\dot\Theta^{\text{obs}}$, approaches zero for $\Gamma \to 0$. In this regard, $\dot\Theta^{\text{obs}}$ behaves like the entropy production rate of the two-dimensional system (no auxiliary variable), where $\dot\Theta^{\text{obs}}=\dot\Theta_1^{\text{obs}}$ is given by Eq.~(\ref{entprod1Cou}).

Increasing $\Gamma$ from zero leads to a monotonic increase of $\dot\Theta^{\text{obs}}$, consistent with our expectation that the larger the external drive, the larger the distance from equilibrium.
In contrast, $\dot\Theta$  behaves non-monotonically, in fact, one observes a (weakly pronounced) minimum at a finite small $\Gamma$. This suggests that shear flow can, to some extent, compensate for the effect of the hidden variable which causes the non-zero value of $\dot\Theta$ at $\Gamma=0$. Clearly, the combined effect of the different sources of nonequilibrium in our model can be highly non-trivial.

We now consider the range around $\Gamma_{\text{crit}}$. Here, we first note that upon approaching the critical shear rate from below,  $\dot{\Theta}$ and $\dot{\Theta}^{\text{obs}}$ do not show any remarkable behavior. This is already surprising:
 From Eq.~(\ref{dotAgom}) it follows that $\dot\Theta$ involves all three components of the angular momenta that are, in turn, linearly related to second moments [see Eq.~(\ref{eq:ell123})].
All of these moments diverge in the limit $\Gamma\to \Gamma_{\text{crit}}$ (see Sec.~\ref{sec:results_stationary} for examples); thus one could indeed expect that $\dot\Theta$ diverges as well.
Moreover, as shown in Fig.~\ref{fig:entrodana} for $\Gamma>\Gamma_{\text{crit}}$, it is possible to obtain analytical results for the entropy production rates even in this transcritical range. 
%
%
To see this, one has to take a closer look at the explicit expressions for the linear combinations of second-order moments appearing in the angular momenta [see Eq.~(\ref{eq:ell123})]
and the entropy production rate. We have calculated these expressions via {\em Mathematica} (results are not shown here). It turns out 
that divergent contributions {\em cancel} in the linear combinations \cite{MAFernandez2023}.
In particular, the total entropy production (for $a=\Omega=1$) given in Eq.~(\ref{dotAgom}) becomes
\begin{equation}
\label{eq:ent_final}
T\dot\Theta=T\dot\Theta_1=\frac{8\left (\Gamma^2-\Gamma+6\right)}{\Gamma+12},
\end{equation}
which clearly does {\em not} diverge at $\Gamma_{\text{crit}}=3$ (note that we restrict ourselves to positive shear rates). Similarly, Eq.~(\ref{dotA1}) for the observed part reduces (for $a=\Omega=1$) to 
\begin{equation}
\label{eq:entobs_final}
T\dot\Theta^{\text{obs}}=T\dot\Theta^{\text{obs}}_1=
  \frac{6\Gamma^2}{12+\Gamma}.
\end{equation}
When using these expressions, we find a smooth continuation of the results in the precritical range, see Fig.~\ref{fig:entrodana}.
A further analytical argument for the continuity of the angular momenta, and thus, the entropy production rate, is given in Appendix~\ref{sec:remarks}.

Of course, there remains the question of the physical meaning
of these transcritical results: at $\Gamma\geq\Gamma_{\text{crit}}$, there is no stable stationary state, which was assumed in the derivation of our expressions. We will come back to this point below.
Finally, we see that, for all shear rates considered, $\dot\Theta_1^{\text{obs}}<\dot\Theta_1$. When considering $\dot\Theta_1^{\text{obs}}$ as a "coarse-grained" result, obtained by focusing only to a subset of variables,
this inequality is indeed expected based on general arguments \cite{PhysRevE.85.041125}.

\begin{figure}
	\centering
\includegraphics[width=0.48\textwidth]{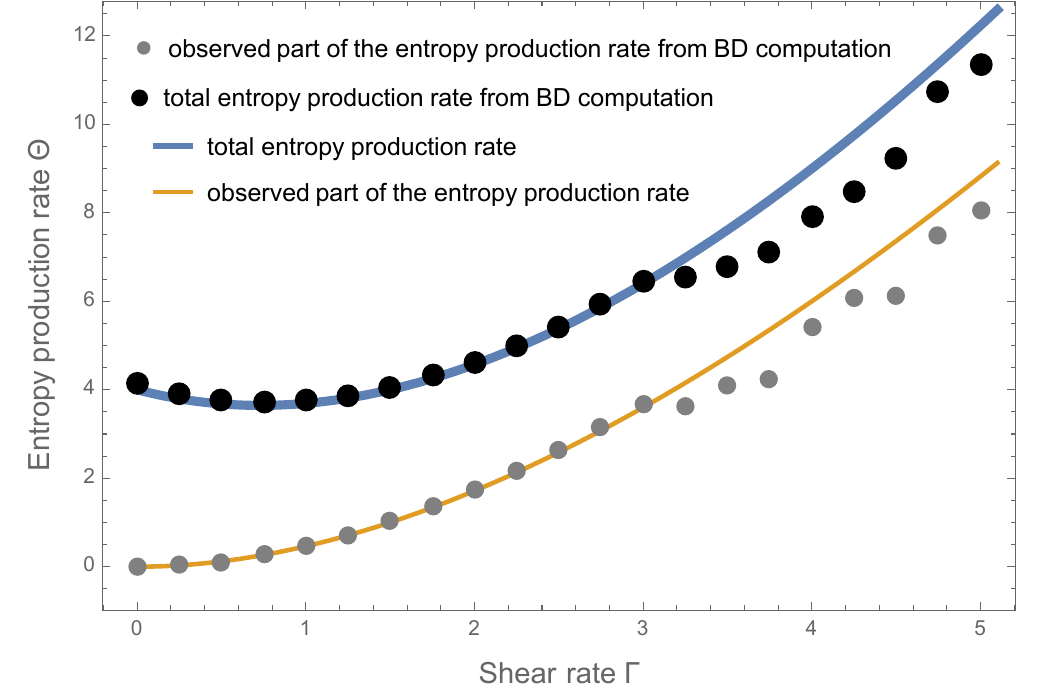}
\caption{The entropy production rate $\dot{\Theta}$ as function of the shear rate $\Gamma$. The dots mark the results from BD simulations  calculated via the angular momenta, while the solid curves
are analytical results from Eqs.~(\ref{eq:ent_final}) and (\ref{eq:entobs_final}.
The upper blue curve and the large black dots pertain to the total entropy production rate,  the lower orange curve and the gray dots  show the  "observed" part of  $\dot{\Theta}$. Parameters as in Fig.~\ref{fig:transR2vsgam}. }
\label{fig:entprodanaBD}
\end{figure}

So far we have studied the analytical results alone.  A comparison with numerical results is presented in Fig.~\ref{fig:entprodanaBD}.  The data have been obtained 
 by direct calculation of the angular momenta  $L_{ik}  = \langle x_i \dot x_k \rangle -   \langle x_k \dot x_i \rangle $ (rather than by the second moments), where $\dot x_i(t)$ is the fluctuating velocity given in Eq.~(\ref{eq:langequzw}).
The large black and the grey dots pertain to the total and to the  "observed" part of  $\dot{\Theta}$, respectively.  The model parameters are the same as for 
 Fig.~\ref{fig:entrodana}.

In the precritical range, $\Gamma<\Gamma_{\text{crit}}$, we find very good agreement between numerical and analytical data, showing the accuracy of the BD simulations. Upon crossing the critical shear rate, the numerical data for $\dot\Theta$ and $\dot\Theta^{\text{obs}}$ behave continuously, and we obtain finite values for the entropy production rates even in the postcritical range. This reflects the performance of the source-sink set-up
in our simulations, or in other words, the fact that we are now considering a finite system with boundaries. As already discussed in Sec.~\ref{sec:results_stationary}, this situation allows the system to reach a quasi-stationary state. Clearly, analytical access is difficult here, since the surface contributions in Eq.~(\ref{def:dotAzw}) [and corresponding relaxation equations (\ref{eq:dtxxbil})] have to be taken into account. 
Numerically, however, we can just compute the entropy production as before.
Interestingly, the so obtained data are surprisingly close to the analytical values from Eqs.~(\ref{eq:ent_final}) and (\ref{eq:entobs_final}), despite the fact that these are, strictly speaking, beyond the limit of applicability. This may be seen as an indirect hint that surface contributions play only a minor role.
 
\begin{figure}
	\centering
\includegraphics[width=0.45\textwidth]{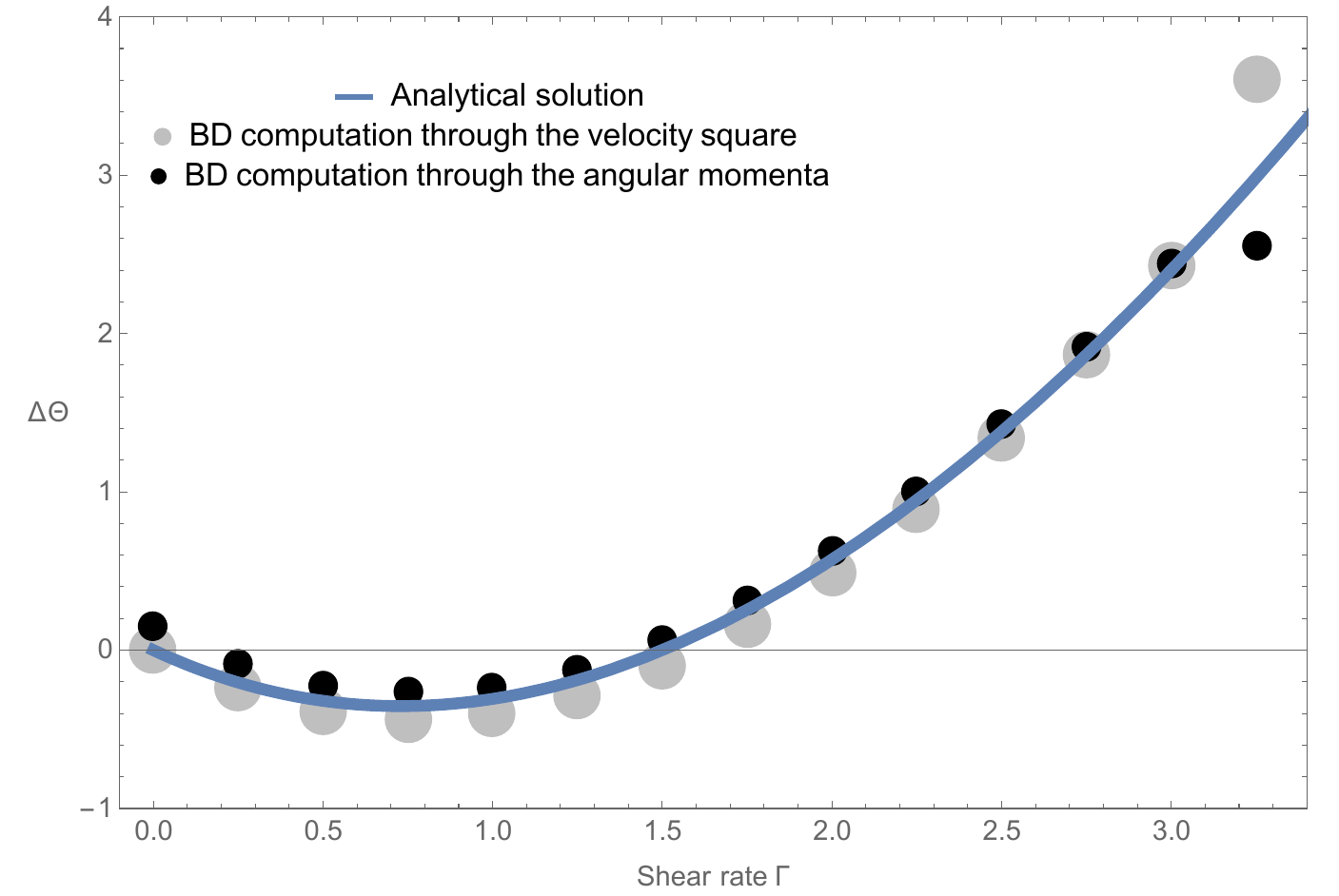}
\caption{Shear flow induced  change of the total entropy production rate, $\delta \dot{\Theta}$, as function of the shear rate $\Gamma$. The black dots represent BD results calculated via the angular momenta; the  curve correspond to the data displayed in the upper part of  Fig.~\ref{fig:entprodanaBD}, shifted downward by the zero-shear value, $\dot\Theta(\Gamma=0)=4$.  The large gray dots stem from BD data computed via the square of the velocity $\dot x_i$. Parameters as in Fig.~\ref{fig:transR2vsgam}.}
\label{fig:entprodanaBDchange}
\end{figure}

 Finally, we have also tested an alternative way to calculate the (total) entropy production rate numerically,
 namely via the square of the velocity, see last line in Eq.~(\ref{entroprod}). In the BD simulations,
$v_i$ has to be replaced by $\dot x_i(t)$, the latter being determined by the right side of Eq.~(\ref{eq:langequzw}). 
%
%
A comparison between the different numerical routes to the total entropy production rate is shown in Fig.~\ref{fig:entprodanaBDchange}. Given that $\dot\Theta$ is nonzero already at $\Gamma=0$
(due to coupling with the hidden variable, see Fig.~\ref{fig:entprodanaBD}), we plot in Fig.~\ref{fig:entprodanaBDchange} directly the difference 
$\Delta \dot{\Theta} \equiv \dot{\Theta} (\Gamma) -   \dot{\Theta} (\Gamma =0) $.
Within computational accuracy, the results obtained via the velocities agree well with those  evaluated  via the angular momenta, 
as presented above. 
\section{\label{sec:concl}Conclusions}
The main purpose of this paper was to explore the impact of memory and nonreciprocal couplings on the dynamical behavior and related thermodynamics 
of a simple model of trapped colloidal particle under shear flow. By specializing on a fully linear model, we were able to obtain analytical results which were then compared with BD simulations.
The memory was induced by coupling of the two physical variables $x$ and $y$ to an auxiliary (hidden) variable, yielding an exponential kernel in a generalized Langevin equation approach. 
Couplings between physical and auxiliary variables have been chosen asymmetrically.
Thus, our model involves several "sources" of nonequilibrium: shear flow (which is nonreciprocal in its own), nonreciprocal coupling with the hidden variable, and the possibility of different heat baths. To deal with the large parameter space, we have mostly focused on the case of uniform temperatures and on couplings where bilinear averages of the two physical variables reduce to equilibrium values in the limit $\Gamma\to 0$.

Regarding the actual impact of the hidden variable on the particle dynamics, our results clearly show that the so-described memory not only affects
quantitatively the shear rate dependence of observable quantities, such as bilinear averages as functions of time and corresponding steady-state values.
Intriguingly, it also gives rise to an instability of the stationary state found for shear rates below a threshold value. In other words, the here considered memory 
leads to a nonequilibrium transition from a stationary into a non-stationary state: the particle escapes the trap.

One might ask to which extent the instability is a result of a specific parameter choice. Are there general conditions for memory-induced instabilities? Within our model, the answer is given by inspection of the determinant
of the deterministic system which is, however, quite involved already in the three-dimensional case [see Eq.~(\ref{eq:detaM})] due to the manifold of parameters. An important point is the flow geometry. For Couette flow,
our results imply that the system without auxiliary variable (i.e., memory) is always stable. 
With auxiliary variable, an instability occurs for the nonreciprocal coupling choices that we have studied here, and that are motivated by requirements concerning the equilibrium limit.
Still, even for reciprocal coupling with the auxiliary variable (e.g., $M_{i3}=M_{3i}$, $i=1,2$)
the Couette system can become unstable. Furthermore, for a different flow geometry, e.g., for extensional flow, an instability can already occur without any auxiliary variable (i.e., memory). 
These examples show that the occurrence of instabilities is indeed a delicate issue.
We also note that memory-induced instabilities are a common feature in systems with time delay \cite{atay2010complex} (where the memory kernel is delta-peaked at a nonzero delay time, corresponding to an infinite number of auxiliary variables \cite{loos2019fokker}) and systems with distributed delay \cite{loos2021medium}.

To further characterize the nonequilibrium behavior of our system, we have explored several thermodynamic notions, with a focus on the ensemble-averaged entropy production rate.
It turns out that this quantity is closely related to a mechanical property, namely the angular momentum (that has also been studied in the context of Brownian 
gyrators \cite{filliger2007brownian,cerasoli2018asymmetry}). For both type of quantities, we have derived exact expressions including the cases of transient dynamics and of boundaries, that are usually neglected.
Further, we have derived explicit steady-state expressions for representative coupling scenarios in dependence of the (instantaneous) second moments and cross correlation between the three variables.

In the absence of the auxiliary variable, the steady-state entropy production rate is nonzero only when the shear flow is turned on, as expected. The auxiliary variable, however, can induce a finite entropy production rate of the full (three-dimensional) system already at $\Gamma=0$. The observed part is, in contrast, zero at $\Gamma=0$ as a consequence of our parameter settings.
A surprising result was that the analytically derived entropy production rates behave continuously when the shear rate crosses the critical value.
Indeed, given that all bilinear averages diverge as the limit of deterministic stability is reached, one would rather expect divergent behavior for the entropy production. 
From an analytical perspective, this ''miracle" is formally explained by the fact that the steady-state entropy production rate is a linear combination of bilinear averages whose divergent contributions cancel out.

The continuity of the entropy production rate beyond the singularity follows from the corresponding behavior of  the angular momentum. Independent from the compensation of diverging terms in the relevant linear combinations, as discussed above, further (analytical) arguments for the observed continuity, based  on properties of  the eigenvalues of the matrix $\mathbf{K} = - \mathbf{a} + \mathbf{M}$, are given in  
Appendix~\ref{sec:remarks}. These  results are specific for our model. On the other hand, non-divergent behavior of the (medium) entropy production rate close to the deterministic limit has also been seen in other linear systems, see \cite{loos2021medium}.
Furthermore, our numerical calculations give another interesting perspective. Within these calculations we have controlled the unstable dynamics beyond the critical shear rate  via a source-sink set-up. It turns out that this allows the system to reach a quasi-stationary state characterized by finite values for the long-time entropy production rate (and bilinear averages).
Moreover, the so-obtained data for $\dot\Theta$ are quite close to the analytical expressions obtained without boundaries.
To summarize, the memory-induced transition is not reflected by the {\em average} values of the entropy production rate and the angular momenta.  Further work in this direction should focus on the distributions and higher-order moments of these quantities, particularly the entropy production rate \cite{MAFernandez2023}. Indeed, test calculations have already shown that, e.g., the fluctuations of ${\ell}_3$ do divergence at the critical shear rate (contrary to ${\ell}_3$ itself). Generally, an investigation of these higher-order moments might also be interesting in the context of fluctuation relations and the thermodynamic uncertainty relation applied to our system, see Ref.~\cite{plati2023thermodynamic} for a related study on another linear system.

Of course, it would be very interesting to see an experimental realization of our results. To this end we first note that the observed instability bears some analogy to the coil-stretch transition of polymer molecules 
\cite{de1974coil,larson1989coil}, where our trapped particle is interpreted as an elastic dumbbell used to model a polymer coil.
With this interpretation, the results obtained here can be used to treat certain rheological properties (such as non-Newtonian viscosities and normal stress differences) of dilute polymer solutions.
Generally, the application of our model of coupled oscillators to
linear polymer molecules, which are often modelled as elastically coupled beads \cite{doi1988theory}, deserves more attention.
For example, it would be interesting to study the dynamics of such a polymer chain under the impact of memory effects modelled by adding auxiliary variables to the "physical" beads.
However, this clearly blows up the parameter space (and, thus, would have been beyond the scope of this article). The same holds if we do not extend the number of physical (observable) degrees of freedom, but rather the number of auxiliary variables, yielding more complex memory (see, e.g., \cite{Doerries_2021}). Still, for the specific case of fully non-reciprocal coupling we do not expect fundamental changes \cite{Loos_2020} (as compared to the case of one hidden variable studied here). 

Regarding corresponding thermodynamics properties, we would like to mention that not the angular momentum (to which the entropy production rate is closely related), but the corresponding angular velocity is, in principle, a measurable quantity 
(see \cite{doi:10.1021/acs.jpcb.3c02324} for a recent example).
Beyond shear-driven polymers, we note that there is recent experimental interest and advances in the (thermo)-dynamics of driven colloids in viscoelastic baths \cite{berner2018oscillating,ginot2022barrier}
and optically trapped colloids subject to a "nonequilibrium" bath with correlated noise \cite{goerlich2022harvesting}, or time-varying temperatures \cite{PhysRevLett.131.097101}. Such systems might be alternative candidates to study memory-induced behavior.

\begin{acknowledgments}
We gratefully acknowledge the support of the Deutsche Forschungsgemeinschaft (DFG, German Research Foundation), project number 163436311 - SFB 910.
\end{acknowledgments}

\appendix
\section{\label{sec:technical}Numerical details}
The BD simulations were conducted using a modified Euler method as described in detail in Ref.~\cite{AAPPAAPP.1011A3,MAFernandez2023}.
The time step in the calculations presented in Figs.~\ref{fig:transR2vsgam} and \ref{fig:transxyqmvsgam} was $\delta t=0.1$.
For the data presented in Figs.~\ref{fig:xvstom1}, \ref{fig:entprodanaBD}, and \ref{fig:entprodanaBDchange}, we used $\delta t=0.04$.
The averages were calculated typically over a run time of $t_{\text{run}} = 1000$, corresponding to $10.000$-$25.000$ time steps
(for each data point corresponding to a specific shear rate).
The numerical accuracy in the precritical range was tested by comparison with exact analytical results.
For the source-sink set-up acting in the deterministically unstable regime, the limiting radius was set to $R_c = 40$. 
When the physical squared distance of the particle from the origin, $r^2=x^2+y^2$ reached $R_c^2$, all three coordinates ($x_1=x$, $y=x_2$, $x_3$) were set back to the origin. 

\section{\label{ssec:bilav}Explicit relaxation equations and stationary solutions}
Here we focus on systems without surface contributions.

The relaxation equations for the first moments (linear averages) have already been discussed below Eq.~(\ref{eq:dtxlin}) and are given by $(d/dt)\langle x_i\rangle=K_{ij}\langle x_j\rangle$ with $i=1,2,3$.
  Inserting the elements of $\mathbf{K}$ and rearranging, one has
   \begin{equation}\label{eq:dynlinav} 
\frac{d}{d t} \langle x_i\rangle    + a^{(i)} \,  \langle x_i\rangle   - M_{i j} \, \langle x_i\rangle = 0 \, .  
   \end{equation} 
In stationary states, $\mathrm{det}(\mathbf{K})<0$ [see Eq.~(\ref{eq:detK})], such that the first moments decay to zero.

We next consider the relaxation equations for two examples of bilinear (second-order) moments, $X_{ij}=\langle x_ix_j\rangle$ (i.e., the elements of the matrix $\mathbf{X}$). 
The time change of the quadratic moment $\langle X_{11}$ follows from Eq.~(\ref{eq:dtxxbil}) as
\begin{equation}
 \label{eq:bilinrel3D}
\frac{1}{2}\frac{d}{d t} X_{11}  +    a^{(1)}   X_{11} -   M_{12}  X_{12}   -  M_{13} X_{13}       =  T^{(1)},
\end{equation} 
while the mixed bilinear moment $X_{12}=\langle xy\rangle$ fullfills 
\begin{eqnarray} 
\label{eq:bilinrel3Dzw}
\frac{d}{d t}   X_{12}  & =& - \left(a^{(1)} +a^{(2)} \right)   X_{12}  +  M_{23}  X_{13}    \nonumber\\   
& & + M_{13} X_{23} +    M_{21}   X_{11}  +  M_{12}  X_{22}.
\end{eqnarray}
The remaining equations follow by cyclic permutation of $1,2,3$. The steady-state solutions can be found by setting the time derivatives to zero. 
A necessary condition for the existence of such a solution is given in Eq.~(\ref{eq:detK}). For $n=3$, the determinant $\text{Det}(\mathbf{K})=D_3$ follows as
\begin{eqnarray} \label{eq:detaM}
D_3 &=& a^{(1)}  a^{(2)}  a^{(3)} -   a^{(1)}  M_{23} M_{32}  - a^{(2)}  M_{31} M_{13} \nonumber\\
& & - a^{(3)}  M_{12} M_{21} \nonumber\\
& & - 
M_{12} M_{23} M_{31} - M_{21}  M_{13}  M_{32}.
\end{eqnarray}
It is understood that all $a^{(i)}  > 0$. 

The linearity of Eqs.~(\ref{eq:bilinrel3D}) and (\ref{eq:bilinrel3Dzw}) allows for analytical calculation of the stationary limits 
of the six quantities $X_{ii}$ ($i=1,2,3$) and $X_{ij}$ ($i \neq j$), depending on the $12$ parameters
$ a^{(i)}$,  $M_{i j }$, and $T^{(i)}$. The latter plays the role of an inhomogeneity in the equations for $X_{ii}$, see Eq.~(\ref{eq:bilinrel3D}). Elimination or rather insertion of the resulting $X_{ii}$ into the equations for 
$X_{ij}$, yields three coupled equations which  can be solved. We note, however, that the actual calculation is quite tedious, we have therefore used Mathematica for the "bookkeeping" as described in \cite{AAPPAAPP.1011A3,MAFernandez2023}.
The resulting expressions are proportional to linear combinations of $T^{(i)}$  and, in general, they are highly  non-linear functions of the remaining nine parameters $ a^{(i)}$ and $M_{i j }$.   
More specifically,   these expressions can be written as 
 \begin{eqnarray}\label{x2HD}
 X_{ii} & = &  \frac{H^{(i i)}}{D_6},\nonumber\\
  X_{ij} &  =& \frac{H^{(ij )}}{D_6}  , \, i \neq j \, , 
 \end{eqnarray} 
where the $H^{(..)}$ and the determinant $D_6$ are polynomials of fifth and sixth degree in the coefficients $ a^{(i)}$ and $M_{i j }$.
The analytic results for arbitrary coupling parameters are presented elsewhere \cite{MAFernandez2023}.
For the case of planar Couette flow [see Eq.~(\ref{MCouette})] and coupling to the auxiliary variable as given in Eq.~\ref{defbirot}) with $\Omega_1=\Omega_2=\Omega$, and for $a^{(i)}=a$, $T^{(i)}=T$, we
find $D_6={\cal D}$ where $\cal{D}$ is given in Eq.~(\ref{eq:avxyCoubirotDet}).


\subsubsection{Second moments for $n=2$}
The situation simplifies in the absence of coupling to the auxiliary variable, i.e., $n=2$. Then, the determinant in Eq.~(\ref{eq:detK}) becomes
\begin{equation}
\label{D2}
D_2=  \left(a^{(1)}  a^{(2)}    -  M_{12} M_{21}\right)  a^{(3)}.
  \end{equation}
For the plane Couette flow ($M_{12} = \Gamma$, $M_{21} = 0$), 
Eq.~(\ref{D2}) implies that stable stationary solutions exist for all values of the imposed shear rate, $\Gamma$. 

The solutions of the moment equations are explicitly given by 
 \begin{eqnarray}\label{mom1122} 
X_{11} &=&  \frac{M_{12} X_{21} + T^{(1)}}{a^{(1)}},\nonumber\\
X_{22} &= & \frac{M_{21} X_{12}  + T^{(2)}}{a^{(2)}},
\end{eqnarray}
and
\begin{eqnarray}
\label{mom12} 
X_{12} &  = & \frac{a^{(1)} a^{(2)}}{D_2\left(a^{(1)}  + a^{(2)}\right)}\nonumber\\
& &\times \left(\frac{M_{12} T^{(2)}}{a^{(2)}}  +  \frac{M_{21} T^{(1)}}{a^{(1)}} \right)
\end{eqnarray}
where $D_2$ is given in Eq.~(\ref{D2}).

\section{\label{sec:angular_general} Angular momentum and torque for linear systems}
Specifying to the linear model at hand (and ignoring surface contributions), where the velocities are given by Eq.~(\ref{eq:velocity}), we find from Eq.~(\ref{def:NL})
\begin{eqnarray}\label{eq:LX}      
L_{ij} &= &    K_{j k} X_{ k i}  - K_{i k} X_{ k j}    \nonumber\\
&=& - ( a^{(j)} -  a^{(i)}) X_{i j} +  M_{j k} X_{ k i}  - M_{i k} X_{ k j},
\end{eqnarray} 
revealing a direct connection between angular momentum and the second moments $X_{ij}$ (i.e., the elements of $\mathbf{X}$) discussed before.
The fluctuating force $F^{\mathrm{fluct}}$ gives no contribution to $L_{ij}$. 
Similarly, using $F_j = K_{j k} x_k  =  - a^{(j)} x_j + M_{j k} x_k $, the torque elements (\ref{def:NL2}) become
\begin{equation}\label{eq:NX}  
N_{ij} =  - (   \gamma^{(j)}   a^{(j)}  -   \gamma^{(i)}   a^{(i)} )  X_{j i}  + 
  \gamma^{(j)} M_{j k}  X_{k i} -   \gamma^{(i)} M_{i k}  X_{k j} \,   .   
\end{equation}
Comparing Eqs.~(\ref{eq:LX}) and (\ref{eq:NX}) we find that, 
in case of isotropic friction (i.e., $\gamma^{(i)}=\gamma$), the elements of $\mathbf{L}$ and $\mathbf{N}$ are proportional to each other, or even identical when we set $\gamma=1$.
This is again a consequence of the overdamped limit and the fact that the fluctuating forces do not contribute to $L_{ij}$. 
\subsubsection{Angular momentum for $n=2$}
From (\ref{eq:LX}) is follows that the angular momentum is determined by
\begin{eqnarray}
\label{eq:LX12dr}      
L_{12}   &=&    \frac{M_{21} T^{(1)}}{ a^{(1)}}   -  \frac{M_{12}  T^{(2)}}{a^{(2)}}   \nonumber\\
& &+
 ( a^{(1)} -  a^{(2)} )  \left(  1 -  \frac{M_{12}  M_{21}}{(a^{(1)}  a^{(2)} } \right)  X_{12},
\end{eqnarray}
where $X_{12}$ is given in Eq.~(\ref{mom12}).
Inserting the relation
\begin{equation}
 \left(  1 -  \frac{M_{12}  M_{21}}{(a^{(1)}  a^{(2)} } \right)  X_{12}
 = \frac{M_{12} T^{(2)}/a^{(2)}  +  M_{21} T^{(1)}/a^{(1)}}{a^{(1)} +  a^{(2)}}
\end{equation}
into Eq.~(\ref{eq:LX12dr})
 we find that  a possible singularity of $X_{12}$ resulting from the determinant $D_2 = 0$ does {\em not} show up in the expression for $L_{12}$. We then obtain, 
 with $a =  ( a^{(1)} +  a^{(2)} )/2$  and  $T =  ( T^{(1)} +  T^{(2)} )/2$, Eq.~(\ref{eq:LX12vi} ) in the main text.

\section{\label{sec:projection}Non-Markovian representation}
In this Appendix we present, as an alternative to the Markovian Langevin equations~(\ref{eq:langequzw}), the corresponding non-Markovian representation  
for the physically observable variables $x_1$ and $x_2$ that results from ``integrating out" the auxiliary variables $x_3$. 
Henceforth we summarize the observable variables within the two-dimensional vector $\mathbf{x}^{o}=(x_1,x_2)$. 

To start with, the solution of the inhomogeneous linear differential equation for $x_3$ is given by
\begin{align}
\label{eq:solution_x3}
x_3(t)&=\int_{t_0}^t dt'\exp[-a^{(3)}(t-t')]\left(M_{31} x_1(t')+M_{32}x_2(t')\right)\nonumber\\
&+\int_{t_0}^t dt'\exp[-a^{(3)}(t-t')]\zeta_3(t').
\end{align}
Henceforth we set $t_0=0$, and $x_3(0)=0$. Inserting Eq.~(\ref{eq:solution_x3}) into the LEs for $x_1$, $x_2$ and rearranging, we obtain the non-Markovian dynamics
\begin{align}
	\dot{\mathbf{x}}^{o}(t) &=\mathbf{K}^{o}\,\mathbf{x}^{o} +\int_{0}^{t}dt'\mathbf{G} (t-t)\,\mathbf{x}^{o}(t')dt'+\bm{\zeta}_{o}+\bm{\zeta}_c.
        \label{eq:x_obs}
\end{align}
In Eq.~(\ref{eq:solution_x3}), $\mathbf{K}^{o}$ is a $2\times2$ matrix containing the elements related to $x_1$, $x_2$ of the coefficient matrix $\mathbf{K}$
introduced in Eq.~(\ref{eq:langequzw}), and $\bm{\zeta}_o$ is the two-component vector of corresponding white noises.
Further, $\mathbf{G}$ is a two-dimensional matrix with elements
\begin{align}
\label{eq:Gij}
G_{ij}(\Delta t)&=M_{i3}M_{3j}\exp\left[-a^{(3)}\Delta t\right],\quad i,j=1,2.
\end{align}
As seen from Eq.~(\ref{eq:x_obs}), $\mathbf{G}(\Delta t)$ plays the role of a kernel that links the dynamics of $\mathbf{x}^o$ to its history. Since we have (only) one linearly coupled auxiliary variable, this kernel involves
one exponential with relaxation time $1/a^{(3)}$.
Finally, the elements of the colored noise $\bm{\zeta}_c$ are given by ($i=1,2$)
\begin{align}
\label{eq:zetaci}
\zeta_{c,i}(t)&=M_{i3}\int_{0}^t dt'\exp[-a^{(3)}(t-t')]\zeta_3(t').
\end{align}
Note that $\zeta_{c,i}$ vanishes if the auxiliary variable is {\em not} coupled to a heat bath, i.e., $T^{(3)}=0$.

Equation~(\ref{eq:x_obs}) can be rewritten by an integration by parts of the second term on the right side. The resulting equation is given by
\begin{align}
\label{eq:gle}
\int_{0}^{t}dt'\mathbf{\Gamma}(t-t')\,\dot{\mathbf{x}}^{o}& = \left(\mathbf{K}^{o}+\frac{1}{a^{(3)}}\mathbf{G}(0)\right)\,\mathbf{x}^o+\bm{\zeta}_{\text{tot}}(t)
\end{align}
with the total noise $\bm{\zeta}_{\text{tot}}=\bm{\zeta}_o+\bm{\zeta}_c$ and the
friction kernel
\begin{align}
\label{eq:friction}
\mathbf{\Gamma}(\Delta t)&=2\mathbf{I}\delta(\Delta t)+\frac{\mathbf{G}(\Delta t)}{a^{(3)}},
\end{align}
where $\mathbf{I}$ is the unity matrix.

Equations~(\ref{eq:x_obs}) and (\ref{eq:gle}) are equivalent, both can be considered as a non-Markovian representation of the LE~(\ref{eq:langequzw}).
Note that although Eq.~(\ref{eq:gle}) is formally similar to a generalized Langevin equation (for an overdamped system), there is an important difference:
here, friction kernel and the noise correlation function are {\em not} automatically linked by a fluctuation-dissipation relation of second kind.

\subsection{Fluctuation-dissipation relation}
Having obtained the non-Markovian representation of our system, it is interesting to investigate whether this fulfills a fluctuation-dissipation relation of second kind (FDR) \cite{RKubo_1966}.
For the present, overdamped system involving two (observable) variables, the FDR can be stated in matrix form as \cite{puglisi2009irreversible,Doerries_2021} 
\begin{align}
\label{eq:FDT1}
\langle\bm{\zeta}_{\text{tot}}(t)\bm{\zeta}_{\text{tot}}(t+\Delta t)\rangle\stackrel{?}{=} \mathbf{T}^o\,\mathbf{\Gamma}(\Delta t),
\end{align}
where $\bm{\zeta}_{\text{tot}}=\bm{\zeta}_o+\bm{\zeta}_c$ is the total noise,
$\mathbf{T}^o$ is the diagonal temperature tensor related to the observable variables (with principal values $T^{(1)}$ and $T^{(2)}$),
and $\mathbf\Gamma$ 
is the (matrix-valued) friction kernel given in Eq.~\ref{eq:friction}).

To inspect the validity of the FDR we have to calculate the two-dimensional matrix of correlation functions of the total noise.
To this end we note that the noise terms for different variables are uncorrelated, therefore
$\langle\bm{\zeta}_o(t)\bm{\zeta}_c(t+\Delta t)\rangle_{ij}=0$,
and $\langle\bm{\zeta}_o(t)\bm{\zeta}_o(t+\Delta t)\rangle_{ij}=2T^{(i)}\delta_{ij}\delta(\Delta t)$. The latter contribution cancels with the delta-like part of the friction kernel Eq.~(\ref{eq:friction}), after multiplying the latter with $\mathbf{T}^o$.

The relation of interest is therefore given by 
\begin{align}
\label{eq:FDT2}
\langle\bm{\zeta}_{c}(t)\bm{\zeta}_{c}(t+\Delta t)\rangle_{ij}\stackrel{?}{=}T^{(i)}\,\frac{G_{ij}(\Delta t)}{a^{(3)}},
\end{align}
where $G_{ij}$ is given in Eq.~(\ref{eq:Gij}) and $i,j=1,2$. 
Using Eq.~(\ref{eq:zetaci}) for $\bm{\zeta}_c$, we find for the elements of matrix of colored noise fluctuations
\begin{align}
\langle\bm{\zeta}_{c}(t)\bm{\zeta}_{c}(t+\Delta t)\rangle_{ij}&=M_{i3}M_{j3}\int_0^{t}du\int_0^{t+\Delta t}dt'\nonumber\\
& \exp\left[-a^{(3)}(t+\Delta t-t')-a^{(3)}(t-u)\right]\nonumber\\
& \times 2T^{(3)}\delta(t'-u),
\end{align}
where the term in the last line results from the correlation $\langle \zeta_3(t')\zeta_3(u)\rangle$. Simplifying we get
\begin{align}
\langle\bm{\zeta}_{c}(t)\bm{\zeta}_{c}(t+\Delta t)\rangle_{ij}&=M_{i3}M_{j3}\exp\left[-a^{(3)}\Delta t\right]\frac{T^{(3)}}{a^{(3)}}\nonumber\\
& \times \left(1-\exp[-2a^{(3)}t]\right).
\end{align}
We now focus on the limit of large $t$. Then, the last exponential in the round brackets can be neglected. Inserting the resulting correlation function into Eq.~(\ref{eq:FDT2}),
using Eq.~(\ref{eq:Gij}), and dividing both sides by $\exp\left[-a^{(3)}\Delta t\right]$,
the FDR  becomes
\begin{align}
\label{eq:FDT3}
M_{i3}M_{j3}\frac{T^{(3)}}{a^{(3)}}\stackrel{?}{=}\frac{T^{(i)}}{a^{(3)}}M_{i3}M_{3j},\quad i=1,2.
\end{align}
From Eq.~(\ref{eq:FDT3}) we see that the FDR is automatically fulfilled if (i) all temperatures are equal and (ii) $M_{j3}=M_{3j}$, that is, the coupling between the observable variables and the hidden variable are reciprocal.
However, if one of these conditions is violated, the FDR can be broken depending on the actual parameter values.
This is indeed the case for the parameter choice in Sec.~\ref{sec:results}, where $T^{(i)}=T$, $i=1,2,3$, but $M_{23}=-M_{32}=\Omega$ and $M_{13}=-M_{31}=-\Omega$.
\section{\label{sec:remarks}Remarks on the continuity of entropy production}
In this Appendix we give additional arguments for the observed continuity of the entropy production and the underlying angular momenta.

We recall that singularities found in stochastic dynamics as treated here are linked with the sign change of the eigenvalues of the matrix $\mathbf{K} = - \mathbf{a} + \mathbf{M}$
appearing in Eqs.~(\ref{eq:langequzw} ) und (\ref{eq:Langevin}).
Thus, the analysis of the solutions of the pertaining deterministic equations sheds light on the stability of observables computed for the stochastic system.

We first consider the deterministic equation $\dot{x_i} = K_{i j} x_j = - a x_i + M_{i j} x_j \, , \,  a > 0$ (resulting from Eq.~(\ref{eq:Langevin}) with zero noise and $a^{(i)}=a$).
Let $\lambda^{(i)}$ be the eigenvalues of the matrix $\mathbf{K}$. The components  $y_i $ in the principal axes system are linked to the original components $x_k$ via an unitarian transformation $y_i = U_{i k} x_k$. Given initial values at $t = 0$, one has 
$y_i \sim \exp[ \lambda^{(i)} \, t]$, for $t > 0$. Solutions of the initial value problem yield $y_i \to   0$ for 
$t  \to  \infty$ provided that  the real part of the eigenvalue obeys $\Re (\lambda^{(i)}) < 0$. 
Under the same condition, the presence of fluctuating forces in the corresponding stochastic equations leads to finite stationary values of the averages $\langle y_i^2\rangle$.   
Unstable solutions  which grow beyond any limits  occur when the real part of  at least  one of the eigenvalues  is positive: $\Re (\lambda^{(i)}) > 0$. 

Now we analyze the special case  $n = 3$ and $\mathbf{M}$ given by Eqs.~\ref{MCouette}) and (\ref{defbirot}).
Here one of the eigenvalues is real, the remaining two are complex conjugate. We use the notation 
$\lambda^{(1 )}  = \lambda$, $\lambda^{(2 )}  = \kappa  + i  \, \nu$, $\lambda^{(3 )}  = \kappa  - i   \, \nu$,
where $\lambda$, $\kappa$ and the eigenfrequency $\nu$ are real quantities.  The principal axes system can be chosen such that the components are real: 
$y_1 \sim \exp[\lambda \,  t ]$, $y_2 \sim \exp[\kappa  \,  t] \,  \cos(\nu t)$, $y_3 \sim \exp[\kappa \,  t] \,  \sin(\nu t)$.

For our completely nonreciprocal parameters, the shear rate dependence of $\lambda $ and $\kappa$ can be inferred from the relations 
 $\Gamma \, \Omega_1 \Omega_2 = (a + \lambda ) ((a + \lambda )^2 + \Omega_1^2 +  \Omega_2^2 )$ and 
 $\lambda + 2 \, \kappa = - 3 \, a$. 
 The first relation follows from Det$(\mathbf{K} - \lambda \, \mathbf{I}) = 0$, the second one is due to the invariance of the trace of $\mathbf{K}$ under unitarian transformations. As expected, one has  $ \lambda =  \kappa = -  a$, for $\Gamma = 0$.  
 For $a = \Omega_1 =  \Omega_2 =1$, the eigenvalue $ \lambda$ turns from negative to positive with increasing shear rate  at $\Gamma =\Gamma_{\text{crit}} = 3$ (see Sec.~\ref{sec:solutions}). At this point, one has $\kappa = - 3/2$. On the other hand, the real part $\kappa$ of the complex pair of eigenvalues changes sign at 
 $\Gamma = - 12$ where $\lambda = - 3$.  This, however, is not relevant here since we restrict our attention to $\Gamma \geq 0$ (where $\kappa < 0$).  
 
The original components $x_i$ obtained by unitarian transformation involve all three principal components. Thus, both exponential factors $\exp{[\lambda t]}$ and $\exp{[\kappa t]}$ determine their time dependence. Consequently, the bilinear products $x_i x_j$ contain linear combinations of $\exp[2 \lambda \, t]$, $\exp[ ( \lambda  + \kappa )\, t]$, and $\exp[ 2 \kappa \, t]$.
 Thus, all bilinear quantities $x_i x_j$  computed for the deterministic system and, as observed above, the averages 
 $\langle x_i x_j\rangle $ computed for the stochastic system, diverge for $\Gamma > 3$ where one has $\lambda  > 0$. 
  
 The situation is different for the components of the angular momentum though they also are expressed in terms of 
 bilinear quantities. Due to 
  $ \dot{y}_1 \sim \lambda \exp[ \lambda   t)]$, $\dot{y}_2 \sim  (\kappa  \cos(\nu t)  - \nu  \sin(\nu t))  \exp{(\kappa t)}  \sim   \exp{(\kappa t)}$, and $\dot{y} _3   \sim  (\kappa \sin(\nu t)  + \nu  \cos(\nu t))  \exp{(\kappa t)}  \sim \exp{(\kappa t)}$,  we have, in our principal axes system 
$\ell_1 = y_2  \dot{y} _3  - y_3  \dot{y} _2  \sim   \exp[ 2\kappa \,  t]$, 
$\ell_2 = y_3  \dot{y} _1  - y_1  \dot{y} _3  \sim   \exp[(\lambda +  \kappa)\,  t ]$, and
$ \ell_3 = y_1  \dot{y} _2 - y_2 \dot{y} _1  \sim   \exp[( \lambda +  \kappa) \,  t ] $.
 The angular momenta in the original coordinate system are linear combinations of these principal components. 
There  is no contribution proportional to $ \exp[2 \lambda \, t]$.  For  $\Gamma > 0$, we have $\kappa < 0$. Further, from the relations above we have $\lambda +  \kappa  = - (3 - \lambda)/2$. This quantity changes sign at $ \lambda = 3$ where 
$ \Gamma = 72$. Thus for $ \Gamma <  72$ also $\lambda +  \kappa < 0$ holds true. To summarize, for the range of shear rates considered in our study, none of the components of the angular momentum diverges. This underlies the  continuous behavior of the average angular momentum  and of the resulting entropy production rate at  the singularity $\Gamma = \Gamma_{\text{crit}} = 3$, as observed for the analytic and numerical calculations presented in the main text.

\bibliography{Fernandez}
\end{document}